\newif\iffigs\figsfalse              
\figstrue                            
\newif\ifbbB\bbBfalse                
\bbBtrue                             

\input harvmac
\overfullrule=0pt

\def\Title#1#2{\rightline{#1}
 \ifx\answ\bigans
  \nopagenumbers\pageno0\vskip1in \baselineskip 15pt plus 1pt minus 1pt
 \else
  \def\listrefs{\footatend\vskip 1in\immediate\closeout\rfile\writestoppt
   \baselineskip=14pt\centerline{{\bf References}}\bigskip{\frenchspacing%
   \parindent=20pt\escapechar=` \input refs.tmp\vfill\eject}\nonfrenchspacing}
  \pageno1\vskip.8in
 \fi 
 \centerline{\titlefont #2}\vskip .5in}

\newcount\figno \figno=0
\iffigs
 \message{If you do not have epsf.tex to include figures,}
 \message{change the option at the top of the tex file.}
 \input epsf
 \def\fig#1#2#3{\par\begingroup\parindent=0pt\leftskip=1cm\rightskip=1cm
  \parindent=0pt \baselineskip=11pt \global\advance\figno by 1 \midinsert
  \epsfxsize=#3 \centerline{\epsfbox{#2}} \vskip 12pt 
  {\bf Fig. \the\figno:} #1\par \endinsert\endgroup\par }
\else
 \def\fig#1#2#3{\global\advance\figno by 1 \vskip .25in
  \centerline{\bf Figure \the\figno} \vskip .25in}
\fi

\ifbbB
 \message{If you do not have msbm blackboard bold fonts,}
 \message{change the option at the top of the tex file.}
 \font\blackboard=msbm10 
 \font\blackboards=msbm7 \font\blackboardss=msbm5
 \newfam\black \textfont\black=\blackboard 
 \scriptfont\black=\blackboards \scriptscriptfont\black=\blackboardss
 \def\Bbb#1{{\fam\black\relax#1}}
\else
 \def\Bbb{\bf}
\fi

\def\NPB#1#2#3{{\sl Nucl. Phys.} \underbar{B#1} (#2) #3}
\def\PLB#1#2#3{{\sl Phys. Lett.} \underbar{#1B} (#2) #3}
\def\PRL#1#2#3{{\sl Phys. Rev. Lett.} \underbar{#1} (#2) #3}
\def\PRD#1#2#3{{\sl Phys. Rev.} \underbar{D#1} (#2) #3}

\def\hepth#1{{\bf hep-th/#1}}
\def\til{\widetilde}
\def\bar{\overline}
\def\bo{\hbox{1\kern -.23em {\rm l}}}
\def\bC{{\Bbb C}}
\def\bR{{\Bbb R}}
\def\bZ{{\Bbb Z}}
\def\tB{{\til B}}
\def\tb{{\til b}}
\def\tQ{{\til Q}}
\def\tq{{\til q}}
\def\tk{{\til\kappa}}
\def\hi{{\hat\imath}}

\def\tnc{{\til n_c}}
\def\stnc{{\tilde n_c}}
\def\t{{}^t\!}
\def\ph#1{\phantom{#1}}

\def\undertext#1{$\underline{\smash{\hbox{#1}}}$}


\Title{hep-th/9603042, RU-96-07, WIS-96-1}
{\vbox{\centerline{The Moduli Space of Vacua of $N=2$ SUSY QCD}
\medskip 
\centerline{and}\medskip
\centerline{Duality in $N=1$ SUSY QCD}}}
\bigskip
\centerline{Philip C. Argyres${}^{1,\star}$, 
M. Ronen Plesser${}^{2,\dagger}$, 
and Nathan Seiberg${}^{1,\ddagger}$}
\smallskip
\centerline{\it ${}^1$Department of Physics and Astronomy, 
Rutgers University, Piscataway NJ 08855 USA}
\centerline{\it ${}^2$Department of Particle Physics, 
Weizmann Institute of Science, 76100 Rehovot Israel}
\centerline{\tt ${}^\star$argyres@physics.rutgers.edu, 
${}^\dagger$ftpleser@wicc.weizmann.ac.il,}
\centerline{\tt ${}^\ddagger$seiberg@physics.rutgers.edu}
\bigskip
\baselineskip 18pt
\noindent
We analyze in detail the moduli space of vacua of $N{=}2$ SUSY QCD with
$n_c$ colors and $n_f$ flavors.  The Coulomb branch has submanifolds
with non-Abelian gauge symmetry.  The massless quarks and
gluons at these vacua are smoothly connected to the underlying
elementary quarks and gluons.  Upon breaking $N{=}2$ by an $N{=}1$
preserving mass term for the adjoint field the theory flows to $N{=}1$
SUSY QCD.  Some of the massless quarks and gluons on the moduli space
of the $N{=}2$ theory become the magnetic quarks and gluons of the
$N{=}1$ theory.  In this way we derive the duality in
$N{=}1$ SUSY QCD by identifying its crucial building blocks---the
magnetic degrees of freedom---using only semiclassical physics and
the non-renormalization theorem.

\Date{3/96}

\newsec{Introduction and Summary}

		\nref\powerd{ N. Seiberg, {\it The Power of
	Duality---Exact Results in 4D SUSY Field Theories,}
	\hepth{9506077}, RU-95-37, IASSNS-HEP-95/46, to appear in the
	Proc.\ of PASCOS '95, the Proc.\ of the Oskar Klein lectures,
	and in the Proc.\ of the Yukawa International Seminar '95.
		} \nref\lectures{ K.  Intriligator and N. Seiberg, {\it
	Lectures on Supersymmetric Gauge Theories and Electric-Magnetic
	Duality,}  \hepth{9509066}, RU-95-48, IASSNS-HEP-95/70, to
	appear in the Proc.\ of Trieste '95 spring school, TASI '95,
	Trieste '95 summer school, and Cargese '95 summer school.
		}
In the last few years it has become clear that supersymmetric field
theories in four dimensions can be analyzed exactly (for a review and a
list of references see \refs{\powerd, \lectures}).  The main new
dynamical insight which has been gained is the role of electric-magnetic
duality.

		\nref\om{C. Montonen and D. Olive, \PLB{72}{1977}{117};
	P. Goddard, J. Nuyts and D. Olive, \NPB{125}{1977}{1}.
		} \nref\dualnf{H. Osborn, \PLB{83}{1979}{321}; A. Sen,
	\hepth{9402032}, \PLB{329}{1994}{217}; C. Vafa and E.  Witten,
	\hepth{9408074}, \NPB{432}{1994}{3}.
		} \nref\SWii{N. Seiberg and E. Witten, \hepth{9408099},
	\NPB{431}{1994}{484}.
		} \nref\SWi{N. Seiberg and E.  Witten, \hepth{9407087},
	\NPB{426}{1994}{19}.
		} \nref\Sei{N. Seiberg, \hepth{9411149},
	\NPB{435}{1995}{129}.
		}
In its simplest form duality is an exact equivalence between
theories \om.  This is the case in the free Abelian theory, in $N{=}4$
SUSY theories \dualnf\ and in finite $N{=}2$ theories \SWii.
Asymptotically free theories are unlikely to exhibit such exact
dualities.  However, if the low energy theory is an Abelian theory, its
duality appears as an ambiguity in the low energy effective
Lagrangian.  This ambiguity in the description plays a crucial role in
solving the theory \SWi.  An alternative notion of duality for
asymptotically free theories was suggested in \Sei\ where two different
dual theories exhibit the same long distance behavior.

		\nref\isson{K. Intriligator and N. Seiberg,
	\hepth{9503179}, \NPB{444}{1995}{125}.
		}
It is important to stress that (with the exception of the duality in
the free Abelian theory) neither the exact Montonen-Olive duality
\om\ nor the duality in $N{=}1$ theories \Sei\ have been proven.  There
is a lot of evidence that these dualities are true.  Furthermore, under
the renormalization group one can flow from a dual pair of $N{=}1$
theories to a dual pair of $N{=}4$ theories \isson, implying that the
duality in $N{=}1$ is a generalization of the duality in $N{=}4$.

		\nref\rlms{R. Leigh and M. Strassler, \hepth{9503121},
	\NPB{447}{95}{1995}.
		}
One of the motivations of this paper is to extend this understanding,
by showing that many of the crucial elements in the duality of \Sei\ in
$N{=}1$ SUSY theories can be traced back to the corresponding $N{=}2$
SUSY theories; for earlier work along these lines, see \rlms.  In
particular, the elementary {\it electric quarks and gluons} of the
$N{=}2$ theory can be continuously deformed to the {\it magnetic quarks
and gluons} of the $N{=}1$ theory.  This explicit demonstration of
their existence is close to a proof of the duality of
\Sei.

		\nref\higgscon{L. Susskind, unpublished; T. Banks, E.
	Rabinovici, \NPB{160}{1979}{349}; E. Fradkin and S. Shenker,
	\PRD{19}{1979}{3682}.
		}
The fact that electric quarks and gluons can be continuously deformed
to magnetic quarks and gluons sounds surprising at first.  However, it
is in accord with the general principles of \higgscon, where it is
shown that one can continuously interpolate between the Higgs and
the confinement phases in theories with matter fields in the
fundamental representation.  Since these two phases are obtained by the
condensation of electric and magnetic charges respectively, one should
be able to continuously deform electric charges to magnetic charges in
such theories.  This was explicitly demonstrated in \SWii\ for Abelian
charges and photons.  Here we extend it to electric and magnetic
non-Abelian quarks and gluons.

We start our analysis in section 2 by studying at the classical level
$N{=}2$ SUSY QCD based on an $SU(n_c)$ gauge theory with $n_f$ quark
hypermultiplets in the fundamental representation.  The moduli space of
classical vacua consists of a Coulomb branch where the gauge group is of
rank $n_c{-}1$ and various Higgs branches where the gauge group is of
lower 
rank.  The different branches touch each other at singular points where
new massless particles are present.  It will turn out to be crucial that
for $n_c \le n_f \le 2n_c {-} 2$ the theory has distinct Higgs branches
touching each other at singular points as shown in Fig.~1.  
		\fig{
Map of the classical moduli space of $N{=}2$ $SU(n_c)$ QCD with
$n_f$ fundamental flavors.  The baryonic and non-baryonic Higgs 
branches intersect along a submanifold $A$, while the non-baryonic
branch intersects the Coulomb branch along submanifold $B$ where 
there is an unbroken $SU(r) {\times} U(1)^{n_c-r}$ gauge symmetry 
with $n_f$ massless fundamental hypermultiplets.  $A$ and $B$ 
intersect at a point where the full $SU(n_c)$ with $n_f$ hypermultiplets 
is unbroken.  There are separate non-baryonic branches for $1 \le r 
\le [n_f/2]$.}{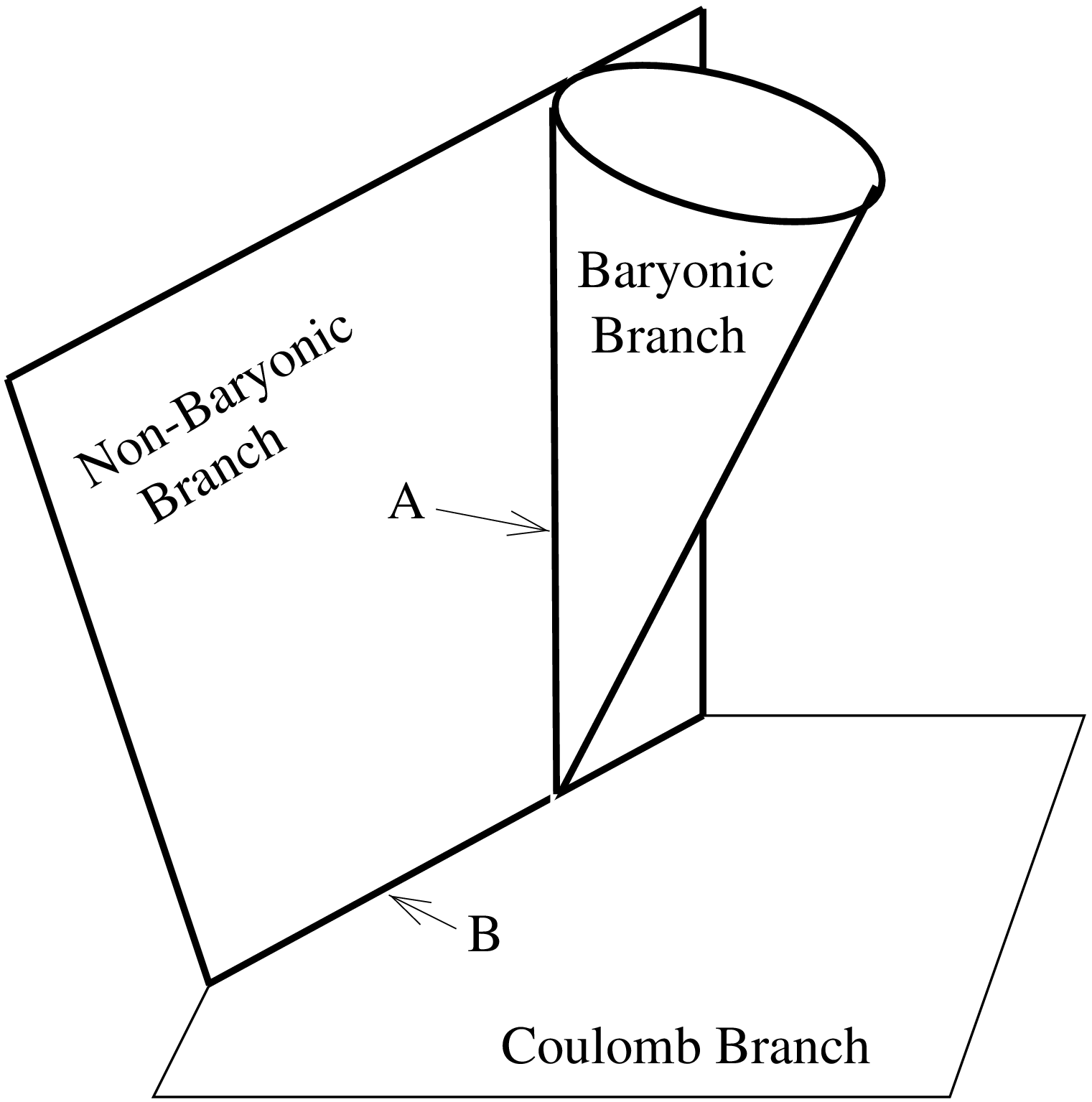}{7cm}

We divide the various Higgs branches into baryonic and non-baryonic
branches, names following from the fact that on the non-baryonic
branches all the light fields have vanishing baryon number.  There will
turn out to be a single baryonic branch for $n_f \ge n_c$ whose generic
low-energy effective theory consists of $n_f n_c {-} n_c^2 {+}1$
massless hypermultiplets.  Non-baryonic branches will be shown to exist
for $n_f \ge 2$, each with (generically) $n_c {-}1 {-}r$ massless
vector multiplets corresponding to a $U(1)^{n_c-1-r}$ low energy gauge
group, and $r(n_f {-}r)$ massless neutral hypermultiplets, where $1 \le
r \le {\rm min}\{[n_f/2], n_c {-}2\}$.  (Only one non-baryonic branch
is shown in Fig.~1 due to lack of dimensions.)

In section 3 we give a very general argument showing
that the Higgs branches as determined classically cannot be modified in
the quantum theory.  This means that the only possible quantum
modification is that distinct Higgs branches which touch classically
may separate, touching the quantum Coulomb branch at different points
due to the splitting of points on the Coulomb branch.  For example,
classically the baryonic and non-baryonic branches meet at the origin
of the Coulomb branch and also intersect along a submanifold.  Since
the origin is split quantum-mechanically, it is possible that the
baryonic branch is split from the non-baryonic branch.  Since, by the
non-renormalization theorem, the Coulomb part of a branch cannot depend
on the squark vevs ({\it i.e.}, on where it attaches to the Higgs
branch), it follows that the baryonic and non-baryonic branches must be
completely disjoint quantum-mechanically (see Fig.~2).  Such a
phenomenon has already been observed in \SWii\ for $n_c=n_f=2$ and here
we demonstrate it for other values of $n_c$ and $n_f$.
		\fig{
Map of the quantum moduli space of $N{=}2$ $SU(n_c)$ QCD with $n_f$
fundamental flavors.  Point $A$ has unbroken gauge group
$SU(n_f{-}n_c)$ with $n_f$ massless fundamental hypermultiplets as well
as various extra monopole singlets.  Submanifold $B$ has unbroken gauge
group $SU(r)\times U(1)^{n_c-r}$ with $n_f$ fundamental hypermultiplets.  
The $X$'s mark points (submanifolds) on $B$ where there are extra
massless singlets.}{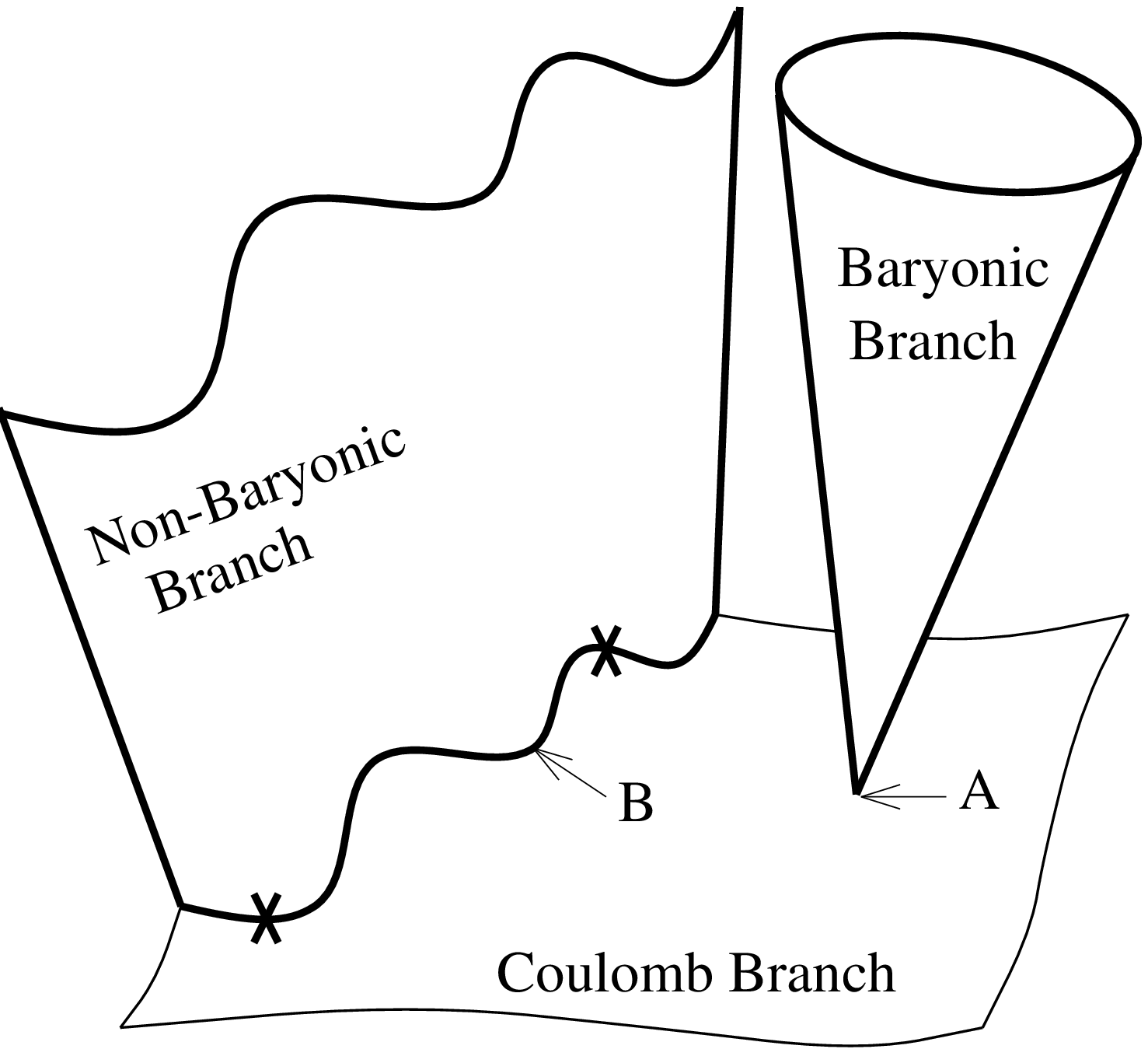}{7cm}

Next we analyze the Coulomb branch at weak coupling.  Semiclassically
the Coulomb branch is characterized by $n_c$ complex numbers (up to
permutation) whose sum vanishes.  These are the eigenvalues $(\phi_1,
..., \phi_{n_c})$ of the complex scalar field $\Phi$ in the adjoint
representation of the gauge group.  At the generic point on the moduli
space where all the eigenvalues are different, the $SU(n_c)$ gauge
symmetry is broken to its Cartan subalgebra $U(1)^{n_c-1}$. The
superpotential $\Tr \tQ \Phi Q$ which couples the adjoint field $\Phi$
to the quarks $Q$ and $\tQ$ makes all the quarks massive at these
points.  On submanifolds where $k$ of the eigenvalues are equal
$SU(n_c)$ is broken to $SU(k) \times U(1)^{n_c-k}$ and the low energy
semi-classical theory includes more gluons.  On submanifolds where some
eigenvalues vanish the classical theory has also $n_f$ massless quarks
(the term in the superpotential $\Tr \tQ \Phi Q$ does not give them a
mass).  Therefore, when $k$ eigenvalues vanish the low energy theory is
$SU(k) \times U(1)^{n_c-k}$ with $n_f$ massless quarks in the
fundamental representation of $SU(k)$.  If $2k < n_f$ the low energy
theory is IR--free and the effective coupling constant of the massless
quarks and gluons vanishes at long distance.  Therefore, the massless
quarks and gluons remain in the spectrum in the full quantum theory.
Similarly, for $2k=n_f$ the low energy theory is finite.  The coupling
constant does not grow at long distance and the massless quarks and
gluons which were found in the classical analysis will remain massless
(though interacting) in the quantum theory.  This semiclassical
analysis is valid for $\Phi = (0,\ldots,0, \phi_{k+1}, \ldots,
\phi_{n_c})$ with all $|\phi_i| \gg \Lambda$.  But given this analysis,
these submanifolds of vacua with massless gluons can be followed in to
strong coupling.

		\nref\AD{P.C. Argyres and M.R. Douglas,
	\hepth{9505062}, \NPB{448}{1995}{93}.
		} \nref\APSW{P.C. Argyres, M.R. Plesser, N. Seiberg,
	and E. Witten, \hepth{9511154}, \NPB{461}{1996}{71}.
		}
We thus conclude that on certain submanifolds of the Coulomb branch the
exact quantum theory must have IR--free massless quarks and gluons or
massless interacting quarks and gluons of a finite $N{=}2$
theory\foot{Non-trivial interacting fixed points such as those of
\refs{\AD,\APSW} can also exist.}.  Some of these will later turn out
to be the magnetic quarks and gluons of the corresponding $N{=}1$
theory.

In section 4 we determine the spectra of massless particles at the
vacua where the various Higgs branches meet the Coulomb branch (the
``roots'' of the Higgs branches).  The physics along these submanifolds
of the Coulomb branch includes the IR--free quarks and gluons described
above.  In the case of the non-baryonic branches the generic theory
at the root is the IR--free or finite $SU(r)\times U(1)^{n_c-r}$ QCD 
with $n_f$ quark hypermultiplets in the fundamental representation and
charged under one of the $U(1)$ factors with $2r \le n_f$.  There
are special points on the submanifolds comprising these non-baryonic roots 
where $n_c {-}r {-}1$ additional singlet hypermultiplets
charged under the $U(1)$ factors become massless (see Fig.~2).  If we
call the singlets $e_i$, then, by an appropriate change of basis
of the $U(1)$'s, their charges can be taken to be
		\eqn\nonbaryonbra{\matrix{
&SU(r)&\times&U(1)_0&\times&U(1)_1&\times&\cdots&\times&U(1)_{n_c-r-1}
&\times&U(1)_B\cr
n_f\times q&\bf r  && 1      && 0      && \cdots && 0      && 0      \cr
e_1        &\bf 1  && 0      && 1      && \cdots && 0      && 0      \cr
\vdots     &\vdots && \vdots && \vdots && \ddots && \vdots && \vdots \cr
e_{n_c-r-1}&\bf 1  && 0      && 0      && \cdots && 1      && 0      \cr
		}}
Just as we have picked a convenient basis for the gauge charges, we
have also used the freedom to shift the global $U(1)_B$ baryon number by
an arbitrary gauge charge.  Our choice is such that the baryon charges of
all the light fields vanish---hence the name of these branches.  There
is also a global $SU(n_f) \times SU(2)_R$ symmetry which is not included
in \nonbaryonbra.

Since the root of the baryonic branch is a single point, it must be
invariant under the discrete global $\bZ_{2n_c-n_f}$ symmetry of the
theory (the anomaly-free part of the classical $U(1)$ R-symmetry).
This suggests that the coordinates of the root on the Coulomb branch
are $\Phi \propto (0,\ldots,0,\omega^1,\ldots,\omega^{2n_c-n_f})$ where
$\omega$ is a $2n_c {-}n_f$-th root of unity.  The $\tnc \equiv n_c {-}
n_f$ zeros imply an unbroken (and IR--free) $SU(\tnc)\times U(1)^{2n_c
- n_f}$ gauge group.  The requirement that there be a Higgs branch
emanating from this vacuum implies that there also be $2n_c {-} n_f$
massless singlet hypermultiplets charged under the $U(1)$ factors.  By
an appropriate change of basis for the $U(1)$'s, the charges can be
taken to be
		\eqn\baryonbra{\matrix{
&SU(\tnc)&\times&U(1)_1&\times&
\cdots&\times&U(1)_{2n_c-n_f}&\times&U(1)_B\cr
n_f\times q&\bf \tnc && 1/\tnc && \cdots && 1/\tnc && -n_c/\tnc \cr
e_1        &\bf 1    && -1     && \cdots && 0      && 0         \cr
\vdots     &\vdots   && \vdots && \ddots && \vdots && \vdots    \cr
e_{2n_c-n_f} &\bf 1  && 0      && \cdots && -1     && 0         \cr
		}}
(Here again we have included the global $U(1)_B$ baryon number charges,
while the $SU(n_f)\times SU(2)_R$ part of the global symmetry was not
included.) We check that \baryonbra\ is the correct identification of
the spectrum at the baryonic branch root by showing that the Higgs
branch emanating from this special vacuum is identical to the baryonic
Higgs branch determined in the classical theory.  As a byproduct, this
argument determines the baryon number of the quarks shown in
\baryonbra.  This spectrum can also be determined by starting with the
finite $n_f=2n_c$ theory and using its conjectured duality to flow down
in flavors by giving masses to quarks.

		\nref\AF{P.C. Argyres and A.E. Faraggi,
	\hepth{9411057}, \PRL{73}{1995}{3931}.
		} \nref\KLTY{A. Klemm, W. Lerche, S. Theisen and S.
	Yankielowicz, \hepth{9411048}, \PLB{344}{1995}{169}.
		} \nref\APS{P.C. Argyres, M.R.  Plesser, and A.D.
	Shapere, \hepth{9505100}, \PRL{75}{1995}{1699}.
		} \nref\HO{A. Hanany and Y. Oz, \hepth{9505075},
	\NPB{452}{1995}{283}.
		} \nref\NM{D. Nemeschanski and J. Minahan,
	\hepth{9507032}, {\it Hyperelliptic Curves for Supersymmetric
	Yang-Mills}.
		} \nref\AS{P.C. Argyres and A.D. Shapere,
	\hepth{9509075}, \NPB{461}{1996}{463}.
		}
In section 5 we review the detailed description of the Coulomb branch
found for the pure gauge $SU(n_c)$ theory in \refs{\AF,\KLTY} and for
the theory with matter in \refs{\APS,\HO}.\foot{The scale-invariant
$SU(3)$ solution found in \NM\ is equivalent to that of \APS, being
related by a reparametrization of the bare coupling---see \AS.}  In the
quantum theory the moduli space is also characterized by $n_c$ complex
numbers $\phi_i$ (up to permutations) whose sum vanishes.  The coupling
constants of the low energy gauge fields form a section of an
$Sp(n_c{-}1,\bZ)$ bundle over the moduli space \SWi\ which was
determined explicitly in \refs{\AF-\HO}.  This allows us to track the
special (singular) submanifolds with enhanced non-Abelian symmetries
found previously into the strong coupling regime.  In particular, we
find the explicit coordinates of the non-baryonic and baryonic roots on
the Coulomb branch.  The monodromies around these singular points
enable us to check the spectrum of massless particles at the
singularities \nonbaryonbra\ and \baryonbra.

In section 6 we use the answers of the previous sections to flow from
$N{=}2$ to $N{=}1$ by giving a mass $\mu$ to the adjoint $N{=}1$
superfield $\Phi$.  Integrating $\Phi$ out of the microscopic theory
we find for $\mu{\gg}\Lambda$, $N{=}1$ $SU(n_c)$ QCD with $n_f$ flavors
describing the theory at scales below $\mu$ but above the
strong-coupling scale of the $N{=}1$ theory, given by a one-loop
matching as 
		\eqn\lambdaone{
\Lambda_1^{3n_c-n_f} = \mu^{n_c}\Lambda^{2n_c-n_f}\ .
		}
To find the extreme low-energy limit, however, we can also first
integrate out the degrees of freedom with mass of order $\Lambda$.
For $\mu{\ll}\Lambda$ this leads to the effective theory described in
the previous sections.  We can then study the breaking to $N{=}1$ as a
deformation of this theory. We find that generic vacua in the moduli
space are lifted by this perturbation, but we show that special points
along the roots of the Higgs branches are not.  Thus the massless
fields in these vacua will descend to the $N{=}1$ theory. (This
analysis cannot rule out the presence of other fields which become
light only in the $\mu{\rightarrow}\infty$ limit.)  The gauge singlets
in \baryonbra\ and
\nonbaryonbra\ and the squarks in \nonbaryonbra\ condense and thus
reduce the gauge symmetry.  However, it is crucial that the squarks in
\baryonbra\ do not condense and therefore the $SU(n_f{-}n_c)$ quarks
and the gluons of equation \baryonbra\ remain massless.  They become
the ``magnetic'' quarks and gluons of \Sei.  Furthermore, in the limit
$\mu {\rightarrow} \infty$ the two branches merge and the light gauge
singlet fields from the the non-baryonic branches contribute to the
extra singlet fields found in the dual $N{=}1$ theory \Sei.  

Thus, we have shown that the extreme low-energy limit of the $N{=}2$
theory is obtained by starting at intermediate scales with either of
the two dual descriptions of \Sei\ (see Fig. 3).  
This is close to a proof of the duality found in that work. 
We note that by explicitly identifying the magnetic degrees of freedom
at an intermediate scale where they are weakly coupled we prove the
exact low energy equivalence (not merely identifying the two chiral
rings) of the two dual $N{=}1$ theories. Moreover, in the $N{=}2$
theory we find that the ``magnetic'' quarks 
and gluons are continuously connected to the ``electric'' degrees of
freedom, by following the singularity associated to their becoming
massless from the weak coupling region in to the strongly-coupled
points where the extra massless gauge singlets appear. 
		\fig{
Renormalization group flows in coupling constant and $\mu$.  The two
trajectories depicted represent the two approaches to the IR fixed
point which approximate the ``electric'' and ``magnetic'' $N{=}1$
theories at intermediate scales.  The diagram holds for $n_f \geq 3/2
n_c$.  For smaller $n_f$ the magnetic $N{=}1$ theory is IR free and
the non-Abelian Coulomb (NAC) fixed point coincides with the 
upper-right-hand corner.  The four corners are labeled by
the gauge group and supersymmetry; all have $n_f$ families of quarks.
The models on the top edge have additional massless singlets as
discussed in the text.
		}{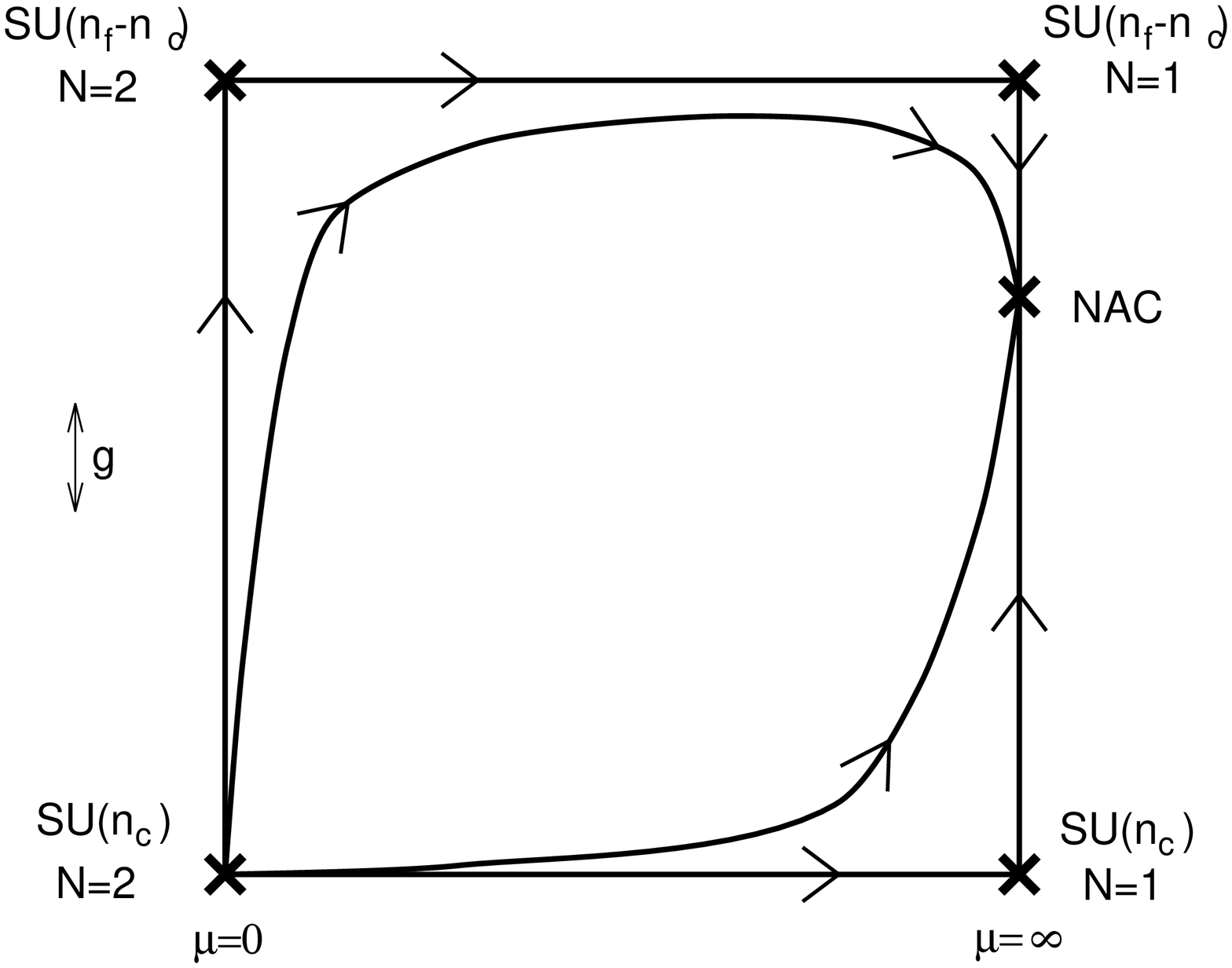}{8cm} 

It is interesting to note how the global symmetries and
quantum numbers are related.   In the $N{=}1$ theory the
global symmetry is $SU(n_f) \times SU(n_f) \times U(1)_B \times
U(1)_R$.  For finite $\mu$ the $U(1)_R$ symmetry is broken and the
$SU(n_f) \times SU(n_f)$ is broken to its diagonal $SU(n_f)$
subgroup.  The extra symmetries of the $\mu{\rightarrow}\infty$ limit
appear as ``accidental'' symmetries of the limiting theory. 
When $\mu {\rightarrow} 0$ the resulting $N{=}2$ theory has global
symmetry $SU(n_f) \times U(1)_B \times SU(2)_R$, where the last factor
is present only for $\mu {=} 0$. 
In relating the two $N{=}1$ theories 
the quarks of the electric theory can switch
their $SU(n_f) \times SU(n_f)$ quantum numbers to become the magnetic
dual quarks since for finite $\mu$ this symmetry is broken to the
diagonal $SU(n_f)$ under which they transform in the same way.  Their
baryon numbers change due to mixing with various gauge $U(1)$ quantum
numbers as we follow the $N{=}2$ vacua in to strong coupling.  In
particular, when $\mu \neq 0$ the various singlets in
\nonbaryonbra\ and \baryonbra\ condense, breaking some of the global
symmetries.  We have defined the $U(1)_B$ in \nonbaryonbra\ and
\baryonbra\ to be the combination of the original baryon number with the
$U(1)$ gauge generators which is left unbroken for non-zero $\mu$.

\newsec{Classical Moduli Space}

\subsec{Symmetries and Vacuum Equations}

		\nref\WB{J. Wess and J. Bagger, {\it Supersymmetry and 
	Supergravity}, second edition, Princeton University, 1992.
		}
$N{=}2$ $SU(n_c)$ supersymmetric QCD is described in terms of $N{=}1$
superfields by a field strength chiral multiplet $W_\alpha$ and a
scalar chiral multiplet $\Phi$ both in the adjoint of the gauge group,
and chiral multiplets $Q^i$ in the $\bf n_c$, and $\tQ_i$ in the ${\bf
\bar n_c}$ representations of the gauge group, where $i=1,\ldots,n_f$
are flavor indices.  $W$ and $\Phi$ together form the $N{=}2$ vector
multiplet, while $Q$ and $\tQ$ make up a hypermultiplet.  We denote the
complex scalar components of $\Phi$, $Q$, and $\tQ$ and their vevs by
the same symbols.  The Lagrangian in terms of $N{=}1$ superfields is
(we follow the conventions of \WB)
                \eqn\Nii{
4\pi{\cal L} = {\rm Im}\Biggl[\tau\int\!\!\! d^2\theta\,d^2\bar\theta\,
{\rm tr}\left(\Phi^\dagger e^V \Phi+ Q^\dagger_i e^V Q^i
+ \tQ^{\dagger i} e^V \tQ_i \right) +
\tau \int\!\!\! d^2\theta\left({\textstyle{1\over2}}{\rm tr}W^2
+{\cal W}\right) \Biggr],
                }
where $\tau$ is the gauge coupling constant and ${\cal W}$ is the
$N{=}2$ superpotential
		\eqn\Niii{
{\cal W}=\sqrt2\tQ^a_i\Phi_a^b Q_b^i+\sqrt2 m^i_j\tQ^a_i Q_a^j,
		}
$a,b=1,\ldots,n_c$ are color indices, and $m^i_j$ is a quark mass
matrix which must satisfy
		\eqn\Vx{
[m,m^\dagger]=0
		}
to preserve $N{=}2$ supersymmetry.  The condition \Vx\ implies that $m$
can be diagonalized by a flavor rotation to a complex diagonal matrix
$m = {\rm diag} (m_1, \ldots, m_{n_f})$.  Classically and with no
masses the global symmetries are a $U(1)_B \times SU(n_f)$ flavor
symmetry and a $U(1)_R \times SU(2)_R$ chiral R-symmetry which acts
on the $N{=}2$ supersymmetry algebra.  Mass terms and instanton 
corrections break $U(1)_R$.  Under the unbroken $SU(2)_R$ the vector 
$A_\mu$ and the scalar $\Phi$ are singlets, while the vector 
multiplet fermions form a doublet.  Similarly for the hypermultiplets,
the fermions are singlets, while their scalar partners $Q$ and
$\tQ^\dagger$ form a doublet.  Though the $SU(2)_R$ action cannot be
made manifest in terms of $N{=}1$ superfields, their transformations
under an R-symmetry $U(1)_J\subset SU(2)_R$ are manifest in \Nii.

When $n_f{<}2n_c$, the theory is asymptotically free and generates a
strong-coupling scale $\Lambda$, the instanton factor is proportional
to $\Lambda^{2n_c-n_f}$, and the $U(1)_R$ symmetry is anomalous, being
broken by instantons down to a discrete $\bZ_{2n_c-n_f}$ symmetry.  For
$n_f=2n_c$ the theory is scale-invariant and the $U(1)_R$ is not
anomalous.  In this case no strong coupling scale is generated, and the
theory is described in terms of its bare coupling $\tau= \theta/\pi +
i8\pi/g^2$.

We describe the selection rules resulting from the breaking of the 
classical symmetries by mass terms and instanton corrections by 
assigning symmetry transformation properties to the corresponding 
parameters in the action.  In particular, the trace of the mass
matrix $m$ is a flavor singlet, $m_S$, while the traceless part
transforms in an adjoint flavor representation, $m_A$.

The symmetry charges for the scalar component fields and parameters are 
thus
                \eqn\Etable{\matrix{
&SU(n_c)&\times&SU(n_f)&\times&U(1)_B&\times&U(1)_R&\times&U(1)_J \cr
Q                  &\bf n_c      &&\bf n_f      &&1  &&0             
&&1 \cr
\tQ                &\bf \bar n_c &&\bf \bar n_f &&-1 &&0             
&&1 \cr
\Phi               &\bf adj      &&\bf 1        &&0  &&2             
&&0 \cr
m_A                &\bf 1        &&\bf adj      &&0  &&2             
&&0 \cr
m_S                &\bf 1        &&\bf 1        &&0  &&2             
&&0 \cr
\Lambda^{2n_c-n_f} &\bf 1        &&\bf 1        &&0  &&2(2n_c{-}n_f) 
&&0 \cr
                }}

The classical vacua are the zeroes of the scalar potential $V =
{1\over2} {\rm Tr} (D^2) + F^{i \dagger} F_i$ where $F_i =
\partial{\cal W} / \partial\varphi^i$, $D^a =
\varphi_i^\dagger(T^a)_j^i \varphi^j$, $\varphi^i$ runs over all
the (dynamical) scalar fields, and $T^a$ are the gauge group
generators in the representation carried by the $\varphi$ fields.  The
classical vacua are thus found by setting the $F$ and $D$ terms to
zero. The $D$-term equations are 
		\eqn\Dvaceqs{\eqalign{ 0 &=
[\Phi,\Phi^\dagger] ,\cr
\nu\delta^b_a &= Q_a^i (Q^\dagger)_i^b - (\tQ^\dagger)_a^i \tQ_i^b ,
		}}
and the $F$-term equations are
                \eqn\Fvaceqs{\eqalign{
\rho\delta^b_a &= Q_a^i\tQ_i^b ,\cr
0 &= Q_a^j m^i_j + \Phi_a^b Q_b^i ,\cr
0 &= m^j_i\tQ_j^a+ \tQ_i^b\Phi_b^a .
                }}
Here $\nu$ and $\rho$ are arbitrary real and complex numbers,
respectively.  These vacuum equations imply that $\Phi$, $Q$ and $\tQ$
may get vevs, which we denote by the same symbols.  {}From the $N{=}1$
point of view, \Dvaceqs\ come from the single $D$-term equation.  The
fact that both equations hold separately
can be seen either by explicitly
squaring the $D$-term and showing that the cross-terms cancel; or by
noting that the first equation is an $SU(2)_R$ singlet while the second
is part of a triplet (with the first equation in \Fvaceqs\ and its
adjoint) so must vanish separately; or, most generally, by noting that
imposing the $D$-term equation and modding-out by gauge transformations
is equivalent to dividing out by {\it complex} gauge transformations
which can be used to diagonalize $\Phi$, thus satisfying the first
equation in \Dvaceqs\ automatically.

We now study the solutions to the vacuum equations \Dvaceqs\ and
\Fvaceqs.  The solutions fall into various ``branches'' corresponding
to different phases of the theory.  The Coulomb phase is defined as the
set of solutions with $Q=\tQ=0$, the (pure) Higgs phase are solutions
with $\Phi=0$, and the mixed phases are those with both nonvanishing
$\Phi$ and $Q$.  For simplicity we will for the most part take
vanishing bare masses $m^j_i =0$.

\subsec{Coulomb Branch}

In the Coulomb phase $\Phi$ satisfies $[\Phi,\Phi^\dagger]=0$ with
$Q=\tQ=0$, implying that $\Phi$ can be diagonalized by a color rotation
to a complex traceless matrix
		\eqn\Ci{
\Phi={\rm diag}(\phi_1,\ldots,\phi_{n_c}), \qquad \sum_a\phi_a=0.
		}
This vev generically breaks the gauge symmetry as $SU(n_c)\rightarrow
U(1)^{n_c-1}$, motivating the name for this branch.

Gauge transformations in the Weyl group of $SU(n_c)$ act on the
$\phi_a$'s by permuting them, so the Coulomb phase is described by
identifying the $n_c{-}1$ complex-dimensional space of $\phi_a$'s under
permutations.  Gauge-invariant coordinates describing the Coulomb
branch can be taken to be a basis of symmetric polynomials in the
$\phi_a$.  This classical moduli space has orbifold singularities along
submanifolds where some of the $\phi_a$'s are equal.  In this case some
of the non-Abelian gauge symmetry is restored.  The scalar potential
also gives the masses of the $Q^i_a$ and $\tQ^a_i$ squarks as
$\phi_a+m_i$.  The vanishing of one of these masses thus describes a
complex codimension one submanifold in the Coulomb phase.

\subsec{Higgs Branches}

The Higgs branch is the space of solutions to the second equation in
\Dvaceqs\ and the first in \Fvaceqs\ since $\Phi=0$.  Describe the
squark fields as $n_c\times n_f$ complex matrices
		\eqn\Hone{
Q = \pmatrix{
Q_1^1     &\cdots &Q_1^{n_f}     \cr 
\vdots    &       &\vdots        \cr 
Q_{n_c}^1 &\cdots &Q_{n_c}^{n_f} \cr} ,\qquad 
\t\tQ =\pmatrix{
\tQ_1^1     &\cdots &\tQ^1_{n_f}     \cr 
\vdots      &       &\vdots          \cr
\tQ^{n_c}_1 &\cdots &\tQ_{n_f}^{n_c} \cr},
		}
where $\t\tQ$ denotes the transpose of $\tQ$.

\medskip
\noindent{{\it 2.3.1}\ 
\undertext{Solutions for the Squark Vevs}}
\medskip

$\bullet$ {\it For} $n_f \ge 2n_c$ any solution of the Higgs branch
equations can be put in the following form by a combination of flavor
and gauge rotations:
		\eqn\Hii{\eqalign{
Q &= \pmatrix{
\kappa_1     &&& 0\ph{{}_1}     &&& 0      &\cr 
&\ddots       && &\ddots         && &\ddots \cr
&&\kappa_{n_c} & &&0\ph{{}_{n_c}} & &       \cr} 
,\qquad \kappa_a\in\bR^+ ,\cr 
\t\tQ &= \pmatrix{
\tk_1        &&& \lambda_1      &&& 0      &\cr 
&\ddots       && &\ddots         && &\ddots \cr
&&\tk_{n_c}    & &&\lambda_{n_c}  & &       \cr}, 
\qquad \lambda_a\in\bR^+ ,\cr
		}}
where
                \eqn\Hiv{\eqalign{
\kappa_a\tk_a &= \rho,\qquad{\rm independent\ of}\ a, 
\qquad \rho\in\bC,\cr
\lambda^2_a &= \kappa_a^2 - {|\rho|^2\over\kappa_a^2}+\nu,
\qquad \nu\in\bR.
                }}
We implicitly assumed above that the $\kappa_a$ were all non-zero.  If
some of the $\kappa_a$ vanish, then the solution \Hii\ and \Hiv\ is
still valid with the proviso that one sets $\rho=0$ before setting any
of the $\kappa_a$'s to zero.

$\bullet$ {\it For} $n_f<2n_c$ solutions can be generated from
$n_f=2n_c$ solutions having corresponding flavor columns of $Q$ and
$\t\tQ$ vanishing, by simply removing the columns in question.
Conversely, to any solution of the Higgs branch equations for $n_f <
2n_c$ one can always add columns of zeros to get a solution with $n_f =
2n_c$, and the flavor rotations necessary to transform the solution
into the form \Hii\ can trivially be chosen not to act on these extra
columns of zeros.  Thus we are assured that this column-reduction
procedure will generate from the $n_f=2n_c$ solutions a solution in
every flavor orbit of the $n_f<2n_c$ solutions.

The only way one can reduce \Hii\ by one or more columns is by setting
$\lambda_1 = \cdots = \lambda_i = \kappa_1 = \cdots = \kappa_j = 0$
with $i {+} j = 2n_c {-} n_f$.  We will find in this way two separate
classes of solutions which we term the baryonic and non-baryonic
branches.  The motivation for this terminology will become clear
shortly.

{\it The Baryonic Branch}.  Choosing $i=2n_c {-} n_f$ and $j=0$, we find
the squark vevs 
                \eqn\bbri{\eqalign{
Q &= \pmatrix{
\kappa_1         &&&&&& \ph{\lambda_1}         &&\cr
&\ddots           &&&&& &\ph\ddots              &\cr 
&&\kappa_{n_f-n_c} &&&& &&\ph{\lambda_{n_f-n_c}} \cr
&&&\kappa_0         &&& &&                       \cr
&&&&\ddots           && &&                       \cr 
&&&&&\kappa_0         & &&                       \cr},
\qquad \kappa_a\in\bR^+ ,\cr
\t\tQ &= \pmatrix{
\tk_1            &&&&&& \lambda_1              &&\cr
&\ddots           &&&&& &\ddots                 &\cr 
&&\tk_{n_f-n_c}    &&&& &&\lambda_{n_f-n_c}      \cr
&&&\tk_0            &&& &&                       \cr
&&&&\ddots           && &&                       \cr 
&&&&&\tk_0            & &&                       \cr},
\lambda_a\in\bR^+ ,\cr
                }}
where
                \eqn\bbrii{\eqalign{
\kappa_a\tk_a &= \rho,\qquad{\rm independent\ of}\ a, 
\qquad \rho\in\bC,\cr
\lambda^2_a &= \kappa_a^2 - \kappa_0^2
+|\rho|^2\left({1\over\kappa_a^2}-{1\over\kappa_0^2}\right).
                }}
We will apply the term baryonic branch to the solutions \Hii\ for $n_f
\ge 2n_c$ as well.  Note that the baryonic branch exists for $n_f \ge
n_c$.  It is not hard to see that reducing \Hii\ as above but with
$i=0$ and $j=2n_c {-} n_f$ leads to a submanifold of the same branch
upon interchanging $Q$ with $\tQ$, which is a symmetry (charge
conjugation) of our theory.\foot{It is not {\it a priori\/} clear that
this symmetry relates points on the same Higgs branch rather than on
two isomorphic Higgs branches.  This follows from the definition, given
below, of what it means for Higgs branches to be separate.}

{\it The Non-Baryonic Branches}.  Another possibility for
column-reduction 
of \Hii\ is to have some of both the $\kappa_a$'s and the $\lambda_a$'s
vanish.  However, from the various constraints in \Hii\ and \Hiv, it
follows that $\rho=\nu=0$ and $\kappa_a=\lambda_a$.  Thus the squark
vevs on this branch are parametrized as
                \eqn\Ii{\eqalign{
Q &= \pmatrix{
\kappa_1 &&& 0\ph{{}_r} &&& 0      &\cr
&\ddots   && &\ddots     && &\ddots \cr
&&\kappa_r & &&0\ph{{}_r} & &       \cr
&&&          &&&            &       \cr
&&&          &&&            &       \cr}, \cr
\qquad \t\tQ &= \pmatrix{
0\ph{{}_r} &&& \kappa_1 &&& 0      &\cr
&\ddots     && &\ddots   && &\ddots \cr
&&0\ph{{}_r} & &&\kappa_r & &       \cr
&&&          &&&            &       \cr
&&&          &&&            &       \cr}, 
\qquad \kappa_a\in\bR^+ ,\cr
                }}
where $r\leq [n_f/2]$ and square brackets denote integer part.  Also,
$2n_c{-}n_f$ columns of zeros should be deleted by the column
reduction procedure.  Note that when $n_f$ is odd, there remains at
least one column of zeros in the reduced matrices.  We will refer to
the manifolds obtained for different values of $r$ as distinct
non-baryonic branches despite the fact that in \Ii\ they appear as
submanifolds of the branch with the maximal value of $r$ by setting
some of the $\kappa_a=0$.  Also, some of the non-baryonic branches are
obtained as submanifolds of the baryonic branch by setting
$\rho=\kappa_0=\tk_0=0$ in
\bbri.  The reason for these distinctions will become clearer below.
Non-baryonic branches exist for $n_f \ge 2$.  For $n_f < 2$
there are no Higgs branches.

\medskip
\noindent{{\it 2.3.2}\ 
\undertext{Gauge Symmetry and Separate Branches}}
\medskip

To make the pattern of intersections of Higgs branches clearer, we
define two Higgs branches to be separate if every path between the two
goes through a point of enhanced gauge symmetry.  In particular, if one
branch has a larger unbroken gauge group than another, they are
separate branches.  This will become clear in subsection 2.4
when we discuss the mixed branches. 

{\it The Baryonic Branch}.  The generic solution \Hii\ or \bbri\
completely breaks the gauge symmetry.  By the Higgs mechanism, the
number of massless hypermultiplets is ${\cal H}= n_f n_c{-}n_c^2{+}1$,
counting the quaternionic dimension of the Higgs branch.  There are
submanifolds of the baryonic branch where the gauge symmetry is
enhanced.  These occur when two or more rows of $Q$ and $\tQ$ vanish.
This can only happen if $\rho=\nu=0$ in \Hii\ or $\rho=\kappa_0=0$ in
\bbri, giving rise to non-baryonic branch vevs \Ii\ with 
		\eqn\restbbr{
r \le {\rm min}\{\, n_f{-}n_c\, ,\, n_c{-}2\, \}.
		}  

{\it The Non-Baryonic Branches}.  For $n_f < 2n_c$ there are
nonbaryonic branches with $r$ outside the range \restbbr . In general
on the nonbaryonic branch the unbroken gauge
group is then $SU(n_c{-}r)$ with $n_f {-} 2r$ massless hypermultiplets
in the fundamental.  Since there are different unbroken gauge
symmetries for each value of $r$ they are separate branches.
(Even though the non-baryonic branch with $r = n_c {-} 1$ does not have
an unbroken gauge symmetry, it is easy to see that it is still separate
{}from the baryonic branch:  any path connecting the two must go through
a point with an unbroken $SU(2)$ gauge symmetry.)  The Higgs mechanism
gives for the number of massless multiplets neutral under the unbroken
gauge group ${\cal H} = r(n_f{-}r)$.

\medskip
\noindent{{\it 2.3.3}\ 
\undertext{Flavor Symmetry}}
\medskip

In order to identify the unbroken global symmetries on the Higgs
branches, it is useful to define a basis of gauge-invariant quantities
made from the squark vevs, the meson and baryon fields
		\eqn\mesbary{\eqalign{
M^i_j &= \tQ^a_j Q_a^i ,\cr
B^{i_1\ldots i_{n_c}} &= Q^{i_1}_{a_1}\cdots Q^{i_{n_c}}_{a_{n_c}}
\epsilon^{a_1\ldots a_{n_c}}\ ,\cr
\tB_{i_1\ldots i_{n_c}} &= \tQ_{i_1}^{a_1}\cdots \tQ_{i_{n_c}}^{a_{n_c}}
\epsilon_{a_1\ldots a_{n_c}}\ .
		}}
The baryon fields are only defined for $n_f \ge n_c$.

{\it The Baryonic Branch}.  On this branch $B,\tB\neq0$, hence its name.  
{}From \Hii\ or \bbri\ the meson field $M^i_j$ is
                \eqn\mesbbr{
M = \pmatrix{
\rho   &&& \kappa_1\lambda_1         &&& 0      &\cr
&\ddots && &\ddots                    && &\ddots \cr
&&\rho   & &&\kappa_{n_c}\lambda_{n_c} & &       \cr
&&&        &&&                           &       \cr 
&&&        &&&                           &       \cr 
&&&        &&&                           &       \cr},
                }
where the $\rho$-block is $n_c\times n_c$.  For $n_f < 2n_c$ one
should remove the appropriate number of columns from the right and
rows from the bottom of \mesbbr.

$\bullet$ {\it For} $n_f \ge 2n_c$, \mesbbr\ and the non-vanishing
baryon vevs imply the global symmetry is broken as\foot{The meson vev
by itself only breaks it to $U(1)^{n_c}$, but the non-vanishing baryon
vevs break baryon number as well.} $SU(n_f)\times U(1)_B\times SU(2)_R
\rightarrow U(n_f {-} 2n_c) \times U(1)^{n_c-1} \times
SU(2)_R^\prime$.  (For $n_f = 2n_c$ drop the $U(0)$ factor.) Thus the
number of (real) Goldstone bosons is ${\cal G} = 4n_fn_c {-} 4n_c^2 {-}
n_c {+} 1$.  Since the number of (real) parameters describing the Higgs
branch in \Hii\ is ${\cal P} = n_c {+} 3$, the fact that ${\cal
G}+{\cal P}=4{\cal H}$ is a check that we have a complete
parametrization of this branch.

$\bullet$ {\it For} $n_c\le n_f<2n_c$ the global symmetry is broken as
$SU(n_f)\times U(1)_B\times SU(2)_R \rightarrow SU(2n_c {-} n_f)\times
U(1)^{n_f-n_c}\times SU(2)_R^\prime$. (For $n_f = 2n_c {-} 1$ drop the
$SU(1)$ factor.)  Thus the number of Goldstone bosons is ${\cal G} =
-4n_c^2 {+} 4n_cn_f {-} n_f {+} n_c {+} 1$, the number of parameters
describing the baryonic branch is ${\cal P} = n_f {-} n_c {+} 3$, and
${\cal G} {+} {\cal P}=4{\cal H}$.

{\it The Non-Baryonic Branches}.
The baryon fields vanish on these branches, $B=\tB=0$, hence their name.
The meson field on these branches is given by
		\eqn\mesnbbr{
M = \pmatrix{
0      &&& \kappa_1^2 &&& 0      &\cr
&\ddots && &\ddots     && &\ddots \cr
&&0      & &&\kappa_r^2 & &       \cr
&&&        &&&            &       \cr 
&&&        &&&            &       \cr 
&&&        &&&            &       \cr},
                }
where the first block of zeros is $r \times r$.  This implies the
global symmetry is broken as $SU(n_f)\times U(1)_B\times SU(2)_R
\rightarrow U(n_f{-}2r) \times U(1)^r\times SU(2)_R^\prime$.  Hence
${\cal G}= r(4n_f {-} 4r {-} 1)$, while ${\cal P} = r$ and ${\cal G}
{+} {\cal P} = 4{\cal H}$.

\medskip
\noindent{{\it 2.3.4}\ 
\undertext{Gauge Invariant Description of the Higgs Branches}}
\medskip

Eq.\ \Hii\ only gives representative solutions for the squark vevs in
the pure Higgs phases.  Global symmetry transformations on these
solutions relate them to inequivalent points in the moduli space
(with identical physics).  Gauge
transformations relate them to equivalent points.  Therefore, it is
useful to describe the moduli space in terms of gauge invariant
coordinates.  Such a description will prove useful in section 4.
So, we would like to describe the various branches in terms of
constraints on the gauge-invariant mesons and baryons \mesbary.

Mathematically, the Higgs branch is a hyperK\"ahler quotient of squark
space by the gauge group, with the $D$- and $F$-terms as moment maps.
We find it useful to work instead with a K\"ahler quotient, essentially
considering the theory as an $N{=}1$ model with a superpotential
interaction.  In a K\"ahler quotient, setting the $D$-terms to zero and
identifying orbits of the gauge group is equivalent to a quotient by
the complexified gauge group.   We achieve this by expressing the vevs
directly in terms of holomorphic gauge-invariant coordinates, the
mesons and baryons \mesbary , and imposing the $F$-term equations.  The
nontrivial structure of the quotient (or the $D$-term equations) is
manifested in the fact that the gauge-invariant coordinates are not
independent as functions of the squarks but satisfy a set of polynomial
relations which we must now impose as constraints.  Below we find a set
of generators for the constraints following from these relations (to
which we refer somewhat loosely as $D$-terms) and the $F$-term
equations.

Since the
product of two color epsilon-tensors is the antisymmetrized sum
of Kronecker deltas, it follows that
		\eqn\mbconstro{
B^{i_1\ldots i_{n_c}}\tB_{j_1\ldots j_{n_c}} =
M^{[i_1}_{j_1} \cdots M^{i_{n_c}]}_{j_{n_c}} ,
		}
where the square brackets imply antisymmetrization.  Also, since any
expression antisymmetrized on $n_c {+} 1$ color indices must vanish, it
follows that any product of $M$'s, $B$'s, and $\tB$'s antisymmetrized
on $n_c {+} 1$ upper or lower flavor indices must vanish.  As long as
both $B$ and $\tB$ are non-zero, an induction argument shows that the
two constraints
		\eqn\mbconstri{
(*B) \tB = *(M^{n_c}) ,
		}
		\eqn\mbconstrii{
M \cdot *B = M \cdot *\tB = 0 ,
		}
imply all the other $D$-term constraints.\foot{A constraint with, say,
$k$ $B$'s and an arbitrary number of $M$'s antisymmetrized on $n_c {+}
1$ upper indices can be replaced by a constraint with $k{-}1$ $B$
fields by \mbconstri.  Repeating this process reduces all constraints
to \mbconstri\ plus the single constraint with no $B$ fields
$*(M^{n_c+1})=0$.  We show below that this latter constraint is implied
by \mbconstrii.} Here the ``$\cdot$'' represents the contraction of an
upper with a lower flavor index, and the ``$*$'' stands for contracting
all flavor indices with the totally antisymmetric tensor on $n_f$
indices.  For example
		\eqn\defstar{
(*B)_{i_{n_c+1} \ldots i_{n_f}} = \epsilon_{i_1 \ldots i_{n_f}} B^{i_1
\ldots i_{n_c}}.
		}
Note that \mbconstri\ is just \mbconstro\ rewritten in this notation.
If $B= \tB=0$, then all the other constraints are automatically
satisfied (being proportional to $B$ or $\tB$) and \mbconstri\ implies
\mbconstrii.\foot{When only one of $B$ or $\tB$ vanishes extra
constraints beyond \mbconstri\ and \mbconstrii\ are required.  However,
such Higgs branches are only submanifolds of the baryonic branch, not
separate branches.  This follows from our explicit parametrization of
the squark vevs \Hii.  Thus we can safely ignore this special case for
the purposes of identifying and counting Higgs branches.} It is useful
to note that \mbconstri\ and \mbconstrii\ imply $0 = \tB (M\cdot *B) =
M \cdot *(M^{n_c}) = *(M^{n_c+1})$, implying
		\eqn\mrankconstr{
{\rm rank}(M) \le n_c .
		}

The $F$-term equations \Fvaceqs\ imply further constraints.  The first
such equation\foot{The other two equations are relevant only for a
description of mixed branches.} implies
		\eqn\mbconstriii{
M' \cdot B = \tB \cdot M' = 0 ,
		}
		\eqn\mbconstriv{
M \cdot M' = 0 ,
		}
as well as another constraint which is just a contraction of
\mbconstri.  In the above we have defined
		\eqn\defmprime{
(M')^i_j \equiv M^i_j - {1\over n_c} (\Tr M) \delta^i_j .
		}
Thus \mbconstri, \mbconstrii, \mbconstriii, and \mbconstriv\ are a 
complete set of constraints.

The $F$-term condition \mbconstriv\ by itself is quite restrictive.
Its only solutions, up to flavor rotations, are in fact precisely
\mesbbr\ and \mesnbbr.  To see this, assume that rank$(M)=r$.  By an
$SU(n_f)$ similarity transformation $M$ can be put into the form
		\eqn\Mblock{
M = \pmatrix{A &B \cr 0 &0 \cr},
		}
where $A$ is an $r\times r$ block and $B$ is an $r\times (n_f {-} r)$
block.  Since rank$(M)=r$, the $r\times n_f$ block $(\,A\,\,B\,)$ must
have $r$ linearly independent columns.  \mbconstriv\ gives $(A -
{1\over n_c}{\rm Tr}A) \cdot (\,A\,\,B\,) = 0$, implying
		\eqn\Ablock{
A^i_j = {1\over n_c} ({\rm Tr}A) \delta^i_j .
		}
The solutions of this equation are either $A$ is diagonal and $r=n_c$,
or $A=0$ and there is no restriction on $r$.  An $SU(n_f)$ similarity
transformation which preserves the form \Mblock\ of $M$ can be chosen
to diagonalize $B$.  Thus the solutions with $r=n_c$ are seen to
coincide with the baryonic branch solutions \mesbbr, while the
solutions with $A=0$ are of the form \mesnbbr\ found for the
non-baryonic branch.  Note that the non-baryonic solutions have rank $r
\le [n_f/2]$.  For $n_f > 2n_c$, this will be reduced to $r \le n_c$ by
\mrankconstr.  For $n_f \le 2n_c$, on the other hand, this constraint
is automatically satisfied, and \mrankconstr\ is implied by
\mbconstriv.

\subsec{Mixed Branches}

So far in our analysis we have not taken the last two equations in
\Fvaceqs\ into account.  They have no effect on the Coulomb branch, and
on the Higgs branches they are clearly satisfied for vanishing masses.
When the masses do not vanish they put constraints on the Higgs vevs.
Intuitively this is clear:  the Higgs phase corresponds to flat
directions along which some components of the quark superfields remain
massless, so bare mass terms tend to lift these flat directions.
Indeed, there are no non-zero masses satisfying the constraint \Vx\ for
which the generic Higgs branch \Hii\ for $n_f=2n_c$ satisfies the last
two equations in \Fvaceqs.

These equations play a more interesting role on mixed branches.  By
mixed branches, we mean solutions to the vacuum equations for which
both $Q$ or $\tQ$ and $\Phi$ are non-zero.  For simplicity assume that
the bare masses vanish.  The $\Phi$ vev can be put in the diagonal form
\Ci\ by a color transformation.  Then the last two equations in
\Fvaceqs\ only have a non-zero solution for $Q$, $\tQ$, and $\Phi$ if
the squark and adjoint-scalar vevs live in disjoint color subgroups.
This leads to a clean distinction between the various Higgs
branches.  Branches with different gauge groups are distinct because
they appear as the Higgs factor of mixed branches with manifestly
distinct Coulomb factors. 

This fact and the other equations in \Dvaceqs\ and \Fvaceqs\ imply that
the vevs can be parametrized up to gauge and flavor rotations as
		\eqn\Iii{
\Phi = {\rm diag}(0,\ldots,0, \phi_{r+1}, \ldots, \phi_{n_c}), \qquad
\phi_a\in\bC , \qquad \sum_a \phi_a =0 ,
		}
and as in \Ii\ for the squarks.  (The squark vevs for $n_f<2n_c$ are
obtained by column reduction of \Ii.)  Thus locally the mixed branch
has the structure of a direct product of a non-baryonic Higgs branch
and a Coulomb branch.  This Coulomb branch can be identified with the
Coulomb branch of the unbroken $SU(n_c{-}r)$ gauge theory of the
non-baryonic Higgs branch.  Henceforth, we will refer to the mixed
branches as non-baryonic branches.

\subsec{Summary of Classical Moduli Space}

This long section has been devoted to solving the classical vacuum
equations of $N{=}2$ supersymmetric $SU(n_c)$ with $n_f$ massless
fundamental flavors.  We can summarize what we have learned by
recording the number ${\cal V}$ of massless $U(1)$ vector multiplets
and $\cal H$ of massless neutral hypermultiplets at a generic point on
the various branches of the moduli space:

\item{$\bullet$} The baryonic branch exists for $n_f \ge n_c$ with
${\cal V} = 0$ and ${\cal H}= n_f n_c - n_c^2 + 1$.

\item{$\bullet$} The non-baryonic branches exist for $n_f \ge 2$ with
${\cal V} = n_c {-} 1 {-} r$ and ${\cal H} = r(n_f {-} r)$, where $1
\le r \le {\rm min} \{\, [n_f/2]\, ,\, n_c {-} 2\, \}$.  (Note that
$[n_f/2]$ is the lesser when $n_f < 2n_c {-} 2$.)

\item{$\bullet$} The Coulomb branch exists for all $n_f$ with ${\cal
V}= n_c{-}1$ and ${\cal H}=0$.

\noindent
We have also learned how the Coulomb, baryonic and non-baryonic
branches fit together.  Classically, Higgs branches intersect the
Coulomb branch at the origin, and out of the Higgs branch emanate
various mixed branches which touch the Coulomb branch along
submanifolds where two or more squarks are massless (which is where
also, classically, a non-Abelian gauge group is unbroken).  Thus,  from
\Iii, when the non-baryonic branches meet the Coulomb branch there is
classically an $SU(r) \times U(1)^{n_c-r}$ unbroken gauge group with
$n_f$ massless hypermultiplets in the fundamental representation of
$SU(r)$ and charged under one of the $U(1)$ factors (by an appropriate
choice of basis of the $U(1)$'s).  Similarly, the baryonic branch meets
the Coulomb branch at its origin where, classically, the full $SU(n_c)$
is unbroken with $n_f$ massless fundamental flavors.  Finally, the
various Higgs branches also connect up along submanifolds of enhanced
gauge symmetry.  In particular, the baryonic branch (with no gauge
symmetry generically) meets the non-baryonic branches on a submanifold
with enhanced gauge group $SU(n_c{-}r)$ for $r \le {\rm min}\{\,
n_f{-}n_c\, ,\, n_c{-}2\, \}$.  This classical picture is sketched in
Fig.~1, where only one non-baryonic branch is depicted.

\newsec{Quantum Higgs Branches and the Non-Renormalization Theorem}
 
So far our analysis has been completely classical.  The question
remains how this structure is modified quantum-mechanically.  We
will now show that there are no quantum
corrections to the baryonic and non-baryonic branches, though where
they intersect each other and the Coulomb branch may be modified.

We have seen above that the classical moduli space is constructed from 
various branches, and the generic point on a given branch is an Abelian
theory with some numbers $\cal V$ of vector multiplets and $\cal H$ of
neutral hypermultiplets.  Far enough out along any of theses branches
the physics is weakly coupled if the microscopic theory is
asymptotically free.  (In the scale-invariant case, $n_f = 2n_c$, we
can make the physics weakly coupled by taking the bare coupling
small.)  Thus in these limits we expect our description of the low
energy effective theory in terms of Abelian vector multiplets and
neutral hypermultiplets to be accurate.  We will now study how it
can change.

		\nref\vPetal{B. de Wit, P.G. Lauwers, and A. Van Proeyen,
	\NPB{255}{1985}{569}.
		}
Consider an arbitrary $N{=}2$ supersymmetric theory of
Abelian vector multiplets and
neutral hypermultiplets.  It was shown in \vPetal\ that the scalar
components of vector multiplets cannot appear in the hypermultiplet
metric, and {\it vice versa}.  For global supersymmetry one can
easily derive that as follows.  Say we have neutral hypermultiplets with
$N{=}1$ chiral multiplets $Q^i$, $\tQ^i$,  $i {=} 1,\ldots,{\cal H}$ and
Abelian vector multiplets containing the $N{=}1$ chiral multiplets
$\Phi^a$, $a {=} 1, \ldots,{\cal V}$.   Denote their scalar components by
$q^i$, $\tq^i$, and $\phi^a$.  $N=1$ supersymmetry implies that the
K\"ahler potential is 
some general function $K(Q^i, \tQ^i, \Phi^a, Q^{i+}, \tQ^{i+},
\Phi^{a+})$.  Then the kinetic terms for the scalars will include the
cross terms $\partial_i \partial_{\overline a} K \cdot \partial_\mu q^i
\partial^\mu \phi^{a+} + c.c.$.  By $N{=}2$ supersymmetry such a term
must be accompanied by a term involving $\partial_i \partial_{\overline
a} K \cdot \overline \psi_{\tq^i} \rlap/\partial \lambda^a$, where
$\psi_{\tq}$ is the hypermultiplet fermion and $\lambda$ the
gaugino.  But such a term is not allowed since to cancel its ($N{=}1$)
supersymmetry variation one needs a term involving two derivatives and
$A^a_\mu$ as well as scalars, out of which no Lorentz-invariant
combination can be formed.  Therefore $\partial_i \partial_{\overline
a}K = 0$, implying $K = K_H(Q^i, \tQ^i, Q^{i+}, \tQ^{i+}) +
K_V(\Phi^a,\Phi^{a+})$.  

This result implies that quantum mechanically
the various branches of the moduli space retain their local product
structure ({\it e.g.} the ``mixed'' non-baryonic branch is still
locally a product of a Coulomb branch and a Higgs branch).

		\nref\nsnrt{N. Seiberg, \hepth{9309335},
	\PLB{318}{1993}{469}.  
		} \nref\strom{A. Strominger, \hepth{9504090},
	\NPB{451}{1995}{96}; K. Becker, M. Becker and A.
	Strominger, \hepth{9507158}, \NPB{456}{1995}{130}.
		}
Further constraints on the quantum theory can be deduced by extending
the non-renormalization theorem of \nsnrt\ to $N=2$ supersymmetry.  We
can promote all the parameters in the theory to background $N=2$
superfields.  The way the parameters appear in the Lagrangian constrain
the supersymmetry representation they belong to.  For example, after
rescaling $Q$ and $\tQ$ in \Nii\ by $\sqrt \tau$ the
gauge coupling constant $\tau$ in \Nii\ multiplies the classical
prepotential.  Hence it can be promoted to a background vector
superfield.  In the quantum theory $\tau$ is replaced by the dynamically
generated scale $\Lambda$ ($\tau \sim \log \Lambda$).  Therefore, we can
think of $\log \Lambda$ as a background $U(1)$ vector superfield.  Since
the metric 
on the Higgs branch is independent of vector superfields, it is
independent of $\Lambda$.  Finally, we can use the fact that the
classical theory is obtained in the limit $\Lambda \rightarrow 0$ to
conclude that that metric is given exactly by the classical
answer.\foot{A similar argument was recently used in string theory
\strom\ to derive a similar non-renormalization theorem.  There the
coupling constant is always a dynamical field -- the dilaton -- and one
does not need to promote a parameter to a background field.}

It is also useful to promote the quark masses $m_j^i$ to background
superfields.  Their couplings in the superpotential \Niii\ are identical
to those of the scalar component of a vector superfield (compare the
first and the second term in \Niii).  These vector superfields
correspond to the gauging of the global $SU(n_f)\times U(1)_B$ flavor
symmetry.  $m_A$ corresponds to the vectors of $SU(n_f)$ while $m_S$ is
associated with $U(1)_B$.  Using this point of view the restriction \Vx\
follows {}from the $D$-term equations for these superfields.  As in the
previous paragraph, we immediately learn that the metric on the Higgs
branch is independent of the masses.

We thus learn that only the Coulomb branch can receive
quantum corrections, and that the mixed non-baryonic branch will retain
its classical product structure of a hypermultiplet manifold times the
vector multiplet manifold corresponding to the subspace of the Coulomb
branch along which the non-baryonic and Coulomb branches intersect.
Though we have not used this fact, the $N{=}2$ supersymmetry implies
that the hypermultiplet and vector multiplet manifolds are further
constrained to be hyperK\"ahler and rigid special K\"ahler manifolds,
respectively.  The rigid special K\"ahler structure has been used to
find the exact Coulomb branch, and this solution will be reviewed in 
section 5.  One result we will need now, though, is that globally the
Coulomb branch can be characterized by $n_c$ complex numbers $\phi_a$
(up to permutations) whose sum vanishes, just as in the classical
analysis of the last section.  These coordinates $\phi_a$ can be viewed
as the eigenvalues of the adjoint scalar $\Phi$ vev.

\newsec{Higgs Branch Roots}

In this and the next section we identify the roots of the baryonic and
non-baryonic branches ({\it i.e.} the submanifolds where they intersect
the Coulomb branch) in the quantum theory.  We will, from now on,
concentrate on the asymptotically free and finite theories with $n_f
\le 2n_c$.  In this section we will identify the physics of the roots
using simple physical arguments.  The upshot of this analysis will be
that the roots of the non-baryonic branches are where the gauge
symmetry is enhanced to $SU(r)\times U(1)^{n_c-r}$ with $n_f$ flavors,
just as in the classical analysis.  The root of the baryonic branch,
though, will be shown to have an unbroken gauge symmetry $SU(n_f {-}
n_c)\times U(1)^{2n_c -n_f}$ with $n_f$ flavors and some singlets.
Note that these are all non-asymptotically-free theories, and so are
weakly coupled in the IR.  In section 5 we use this information to
precisely locate the roots using the exact solution for the Coulomb
branch.

\subsec{The Non-Baryonic Root}

In section 2 we identified the non-baryonic root as the submanifold of
the Coulomb branch with unbroken $SU(r)\times U(1)^{n_c-r}$ gauge
symmetry for $r \le [n_f/2]$, with $n_f$ massless fundamental
hypermultiplets charged under a single $U(1)$ factor.  When $r <
[n_f/2]$, or $r = [n_f/2]$ and $n_f$ is odd, this gauge theory is
IR--free, and thus we expect that the non-Abelian factor will not be
broken in the infrared by quantum-mechanical effects.  When $r =
[n_f/2]$ and $n_f$ is even, the non-Abelian factor is scale invariant.
{}From the exact solution for the low-energy effective action on the
Coulomb branch of such theories it is known 
that a vacuum with unbroken $SU(n_f/2)$ gauge symmetry
survives in the infrared
(this is shown in section 5).  We thus conclude that even with
quantum-mechanical corrections, the non-baryonic root will remain the
submanifold of unbroken gauge group determined classically.

We can make a few checks on this conclusion.  Since the theory at the
root is weakly coupled, we can use it to compute the Higgs branch
emanating from it.  By the nonrenormalization theorem the result
should be the same as the classical computation of the non-baryonic
branches performed in section 2.  Recall that that computation found
the non-baryonic branch has ${\cal V} = n_c {-} r {-} 1$ massless
$U(1)$ vector multiplets.  These clearly correspond to the $U(1)$
factors of the IR theory under which none of the hypermultiplets are
charged.  In section 2 we also learned that the non-baryonic branch has
${\cal H} = r(n_f {-} r)$ massless neutral hypermultiplets, an unbroken
global symmetry group of $U(n_f {-} 2r) \times U(1)^r \times SU(2)_R$,
and can be described by squark vevs of the form \Ii.  These properties
can be easily recovered from the IR theory.  The relevant gauge group
(under which the hypermultiplets are charged) is $SU(r) \times U(1)$,
so the analysis of the Higgs branch is similar to that of section 2
with the number of colors now given by $r$.  The condition coming from
the $U(1)$ factors amounts to setting $\nu = \rho = 0$ in \Dvaceqs\ and
\Fvaceqs.  The $n_f \ge 2n_c$ squark vevs \Hii\ then indeed take the
form \Ii, and it follows that the number of hypermultiplets and global
symmetries match those of the non-baryonic branch.

Along the root of the non-baryonic branch, though generically there is
no light matter charged under the $n_c {-} r {-} 1$ $U(1)$'s, it is
found from the exact solution for the Coulomb branch that there are
special submanifolds where some (monopole) hypermultiplets charged with
respect to these $U(1)$'s become massless.  Such special vacua will be
important in breaking to $N{=}1$ supersymmetry (section 6) which lifts
all the vacua in the non-baryonic root except for those with a massless
charged hypermultiplet for each $U(1)$ factor.  Such vacua occur at
isolated points on the baryonic root---we will determine their
coordinates in section 5 using the exact solution on the Coulomb
branch.  We will find that, by a combination of electric-magnetic
duality rotations in the $U(1)$ factors and an appropriate change of
basis among the $U(1)$'s, we can take the singlet hypermultiplets to
each have charge 1 under a single $U(1)$:
		\eqn\Enonbary{\matrix{
&SU(r)&\times&U(1)_0&\times&U(1)_1&\times&\cdots&\times&U(1)_{n_c-r-1}
&\times&U(1)_B\cr
n_f\times q&\bf r  && 1      && 0      && \cdots && 0      && 0      \cr
e_1        &\bf 1  && 0      && 1      && \cdots && 0      && 0      \cr
\vdots     &\vdots && \vdots && \vdots && \ddots && \vdots && \vdots \cr
e_{n_c-r-1}&\bf 1  && 0      && 0      && \cdots && 1      && 0      \cr
		}}
Here $q^i$ are the $n_f$ quark multiplets, and the $e_k$ label the $n_c
{-} r {-} 1$ singlet hypermultiplets at such a special point.  
We have included the $U(1)_B$ global baryon numbers.
Since we are free to redefine a global charge by adding arbitrary
multiples of any local charges, it is clear from \Enonbary\ that we are
able to set the baryon number of all the fields to zero as shown.
We find $2n_c {-} n_f$ such vacua, related by a
$\bZ_{2n_c-n_f}$ discrete symmetry on the Coulomb branch (the 
anomaly-free subgroup of the classical $U(1)_R$ symmetry) which leaves 
the non-baryonic root invariant. (The case $r=n_f-n_c$ is an exception
to this last statement.)

\subsec{The Baryonic Root}

Locating the baryonic root is more difficult.  The reason is that
classically the baryonic root is the origin of the Coulomb branch, where
the full $SU(n_c)$ gauge group is unbroken.  But, for $n_f < 2n_c$ this
vacuum is asymptotically-free (AF), and thus will be altered in the IR by
strong quantum effects.  Indeed, in the exact solution for the Coulomb
branch, it is found that the singularity corresponding to the classical
vacuum does not appear, and instead is ``split'' into various
singularities all at distances $\sim \Lambda$ from the origin.  Here
$\Lambda$ is the dynamically generated strong-coupling scale of the AF
theory.

The one exception to this strong-coupling difficulty is the
scale-invariant case, when $n_f = 2n_c$, and the bare coupling, $\tau$, is
a parameter of the theory.  At weak coupling, $\tau \rightarrow +i\infty$,
we know that the vacuum in question is at the origin of the Coulomb
branch.  (Using the exact solution, we learn in section 5 that it remains 
there even at strong coupling.)

For $n_f < 2n_c$, the $U(1)_R$ symmetry of the classical theory is
broken by anomalies to a discrete $\bZ_{2n_c - n_f}$ symmetry.  This
acts on the $\Phi$ vev by multiplication by a phase.  Since the
baryonic root is a single point (a fact which cannot be modified
quantum-mechanically by the non-renormalization theorem) it must be
invariant under the $\bZ_{2n_c-n_f}$ symmetry.  For $n_f \ge n_c$ the
most general diagonal $\Phi$ vev (coordinates on the Coulomb branch)
which are invariant are of the form\foot{Larger-dimensional submanifolds
with this property exist for $n_f \ge {3\over2}n_c$:  The $k$-complex
dimensional submanifold
		$$
\Phi = (0\,,\,\ldots\,,\,0,
\underbrace{\varphi_1 \omega^i}_{2n_c-n_f},\,\ldots\,, 
\underbrace{\varphi_k \omega^i}_{2n_c-n_f})
		$$
exists for $n_f \ge ({k+1 \over 2k}) n_f$, along which the unbroken
gauge group is $SU(kn_f - (2k{-}1) n_c) \times U(1)^{2kn_c - kn_f}$.
It will turn out, however, that there are no points along this
submanifold where $2kn_c {-} kn_f$ or more monopoles become massless;
see section 5.}
		\eqn\bbrcoord{
\Phi \propto (\underbrace{0, \ldots, 0}_{n_f-n_c} ,
\omega,\omega^2,\ldots,\omega^{2n_c-n_f}),
		}
where $\omega = {\rm exp}\{2\pi i/(2n_c - n_f)\}$.  Note that, up to
overall normalization, there is only one such point since the
components of $\Phi$ are identified up to permutations.  Points on the
submanifold \bbrcoord\ classically have unbroken gauge group $SU(n_f
{-} n_c) \times U(1)^{2n_c - n_f}$ with $n_f$ fundamental
hypermultiplets.  Since this is an IR--free theory, we expect its gauge
symmetry to survive quantum-mechanically.  Quantum effects could,
however, change this effective theory by bringing down additional light
degrees of freedom.  In particular, there may be points on the
submanifold \bbrcoord\ where (monopole) singlets charged under the
$U(1)$ factors become massless.  It is straight-forward to show that
such a theory has no purely hypermultiplet Higgs branch (like the
baryonic one) unless there is at least one singlet charged under each
$U(1)$ factor.  Let us assume there is such a point with precisely
$2n_c {-} n_f$ singlets; in section 5 we will show that such a point
exists and compute the normalization of $\Phi$ in \bbrcoord\ for this
point using the exact solution.

If the effective theory at the baryonic root is $SU(n_f {-} n_c)\times
U(1)^{2n_c-n_f}$ with $n_f$ massless squarks $q^i$ and $2n_c{-}n_f$
massless singlets $e_k$, we can
pick a basis in which the singlets have a diagonal charge
matrix, with each singlet having charge $-1$ under only one of the
$U(1)$'s.  We will show later that the squarks then have charge $1/(n_f
{-} n_c)$ under each of the
$U(1)$ factors.\foot{This is determined later in this subsection by the
requirement that the Higgs branch be correctly reproduced.  In the
next subsection we will rederive this result from a different argument.}
These charges can be summarized as follows:
		\eqn\macbary{\matrix{
&SU(\tnc)&\times&U(1)_1&\times&\cdots&\times&U(1)_{n_c-\tnc}&
\times&U(1)_B\cr
n_f\times q  &\bf \tnc && 1/\tnc && \cdots && 1/\tnc && -n_c/\tnc \cr
e_1          &\bf 1    && -1     && \cdots && 0      && 0         \cr
\vdots       &\vdots   && \vdots && \ddots && \vdots && \vdots    \cr
e_{n_c-\tnc} &\bf 1    && 0      && \cdots && -1     && 0         \cr
		}}
where we have defined
		\eqn\ntil{
\tnc \equiv n_f - n_c ,
		}
a combination which will appear frequently below.  We have included the
charges under the global $U(1)_B$ baryon number.  An appropriate
combination of any such $U(1)_B$ with the gauge $U(1)$'s will eliminate
the baryon number of the singlets, as shown.  The normalization of the
baryon number will be determined below by matching to the elementary
squarks $Q$ whose baryon number is defined to be 1.

We now would like to show that the vacua \macbary\ are indeed the root
of the baryonic Higgs branch.  First of all, this is plausible because
the dimensions and global symmetries of the two Higgs branches are the
same.  To see this, write down the vacuum equations for the theory
\macbary.  Let $\phi$ be the adjoint scalar component of the vector
multiplet for the $SU(\tnc)$ gauge group, and $\psi_k$, $k= 1, \ldots,
n_c {-} \tnc$ be the adjoint scalars for the $U(1)$ multiplets.  Then
the superpotential for the theory is
		\eqn\bbraspot{
{\cal W}/\sqrt2 = 
\tr(\tq\cdot q\phi) + {1\over\tnc} \tr(q\cdot\tq) \Bigl(\sum_k
\psi_k\Bigr) - \sum_k \psi_k e_k \til e_k,
		}
where the trace is over the $\tnc$ color indices and the dot product is a
sum over the $n_f$ flavor indices.  The resulting $D$-term equations are
		\eqn\bbrD{\eqalign{
|e_k|^2 - |\til e_k|^2 &= \nu,\cr
q \cdot q^\dagger - \tq^\dagger\cdot \tq &= \nu, \cr
[\phi, \phi^\dagger ]&= 0 ,
		}}
and $F$-term equations are
		\eqn\bbrF{\eqalign{
e_k \til e_k &= \rho ,\cr
q \cdot \tq &= \rho, \cr
\psi_k e_k &= \psi_k \til e_k  = 0, \cr
\Bigl(\tnc\phi + \sum_k \psi_k\Bigr)q &=
\tq \Bigl(\tnc \phi + \sum_k \psi_k\Bigl) = 0 ,
		}}
with no summation implied over repeated
indices, and with $\nu \in \bR$, $\rho \in \bC$.  The $\phi$ and $q$
equations are essentially the same as in \Dvaceqs\ and \Fvaceqs.  On
the pure Higgs branch ({\it i.e.} setting $\psi_k = \phi = 0$) these
equations determine $e_k$ and $\til e_k$ up to phases, and the
remaining equations for $q$ and $\tq$ are the same as the Higgs branch
equations solved in section 2.3.  Thus, the generic solution up to
flavor rotations is given by \Hii\ and \Hiv\ with $n_c$ replaced by
$\tnc$.  This solution completely breaks the $SU(\tnc) \times
U(1)^{n_c-\tnc}$ gauge symmetry.  Since the total number of
hypermultiplets is $n_f\tnc {+} n_c {-} \tnc$, by the Higgs mechanism
the number of massless hypermultiplets remaining on this branch is
${\cal H} = n_f\tnc {+} n_c {-} \tnc {-} (\tnc^2 {-} 1 {+} n_c {-}
\tnc) = n_f n_c {-} n_c^2 {+} 1$, matching that found in section 2.3
for the 
baryonic branch.  Furthermore, it is easy to see that the global
symmetry is also broken in the same way: $SU(n_f) \times U(1)_B
\rightarrow U(n_f{-} 2\tnc) \times U(1)^{\tnc-1} = SU(2n_c {-} n_f)
\times U(1)^\tnc$.

By the non-renormalization theorem, the Higgs branch of the effective
theory at the root of the baryonic branch \macbary\ should be precisely
the same as the baryonic branch of the electric theory.   We will now
show this by describing the Higgs branches of the effective theory
\macbary\ by equations for its gauge-invariant fields and then finding
a change of variables mapping them to the meson and baryon constraints
\mbconstri, \mbconstrii, \mbconstriii, and \mbconstriv\ derived in 
section 2.3.4.

The $D$-term vacuum equations \bbrD\ are solved by writing all the 
holomorphic invariants of the complexified gauge group.  An overcomplete
basis of such invariant fields is
		\eqn\bbrbasis{\eqalign{
N^j_i &\equiv q^j \tq_i, \cr
E_k &\equiv e_k \til e_k, \cr
b^{i_1 \ldots i_\stnc} &\equiv q^{i_1}_{a_1} \cdots q^{i_\stnc}_{a_\stnc}
\epsilon^{a_1 \ldots a_\stnc}\,\prod_{k=1}^{n_c-\tnc} e_k, \cr
\tb_{i_1\ldots i_\stnc}&\equiv\tq_{i_1}^{a_1}\cdots\tq_{i_\stnc}^{a_\stnc}
\epsilon_{a_1 \ldots a_\stnc}\,\prod_{k=1}^{n_c-\tnc} \til e_k.
		}}	
Note that gauge invariance of these requires the $U(1)$ charges of $q$
to be those of \macbary .
The first two $F$-term equations in \bbrF\ imply
		\eqn\bbrEtoN{
E_k = {1\over \tnc} {\rm Tr}N .
		}
The constraints satisfied by $N$, $b$, and $\tb$, by virtue of
their definitions \bbrbasis\ are of the same nature as those satisfied
by the $M$, $B$, and $\tB$ fields discussed in section 2.3.4. 
By similar arguments their $D$-term constraints can be reduced to
		\eqn\bbrconstri{
(*b)\, \tb = \left({1\over\tnc}{\rm Tr}N\right)^{n_c-\tnc}\ *(N^\tnc),
		}
		\eqn\bbrconstrii{
N \cdot *b = N \cdot *\tb = 0,
		}
when neither $b$ nor $\tb$ (nor therefore Tr$N$) vanishes.  In this
case note that \bbrconstri\ and \bbrconstrii\ imply
		\eqn\nrankconstr{
{\rm rank}(N) \le \tnc.
		}
When both $b=\tb=0$, all the $D$-term constraints except
\bbrconstri\ and \nrankconstr\ become trivial.  Note that, unlike the
case for the microscopic constraints found in section 2.3.4,
\bbrconstri\ does {\it not} imply \nrankconstr\ when $b=\tb=0$, and
thus \nrankconstr\ must be added to the list of independent constraints
(though it is only independent for the non-baryonic branches).\foot{The
case when only one of $b$ and $\tb$ vanishes requires a separate
argument.  As noted in section 2.3.4, there are more constraints for
this case; however, their solutions are just submanifolds of the
baryonic branch, as can be seen from explicit solutions of \bbrD\ and
\bbrF\ for the squark vevs.}

The remaining $F$-term equations in \bbrF\ imply
		\eqn\bbrconstriii{
N' \cdot b = N' \cdot \tb = 0,
		}
		\eqn\bbrconstriv{
N \cdot N' = 0 ,
		}
as well as another constraint which is just a contraction of \bbrconstri.
We have defined 
		\eqn\defnprime{
(N')^i_j \equiv N^i_j - {1\over \tnc} ({\rm Tr} N) \delta^i_j .
		}
Thus, \bbrconstri--\bbrconstriv\ form a complete set of constraints
for the Higgs branches of the IR--effective theory \macbary.

We can map the constraints for the microscopic theory into the
constraints of the IR--effective theory by the change of variables
		\eqn\mapping{
M = N',\qquad B = (-)^{n_c}\, *\tb ,\qquad \tB = (-)^{n_c}\, *b ,
		}
whose inverse is
		\eqn\mappinv{
N = M',\qquad b = (-)^{\tnc}\, *\tB ,\qquad \tb = (-)^{\tnc}\, *B .
		}
Note that the mapping \mapping\ implies the normalization of the baryon
number given in \macbary.  To prove the claim that the branches of the
IR--effective theory \macbary\ are a subset of the branches of the
original microscopic $SU(n_c)$ theory, we must show that
\bbrconstri--\bbrconstriv\ imply \mbconstri, \mbconstrii, \mbconstriii,
and \mbconstriv.

It follows immediately from \mapping\ and \mappinv\ that
\bbrconstriv\ is equivalent to \mbconstriv, \bbrconstrii\ to
\mbconstriii, and \bbrconstriii\ to \mbconstrii.
Thus, we only need to show that \mbconstri\ is satisfied.  On a branch
with both $b$ and $\tb$ non-zero, we must have, by \bbrconstri, Tr$N
\neq 0$, implying Tr$M \neq 0$.  {}From section 2.3.4, the solution to
\mbconstriv\ with Tr$M \neq 0$ is
		\eqn\Mbb{
M = \pmatrix{
\rho &       &       &     &\lambda_1 &       &             \cr
     &\ddots &       &     &          &\ddots &             \cr
     &       &\ddots &     &          &       &\lambda_\tnc \cr
     &       &       &\rho &0         &\cdots &0            \cr
\ph0 &       &       &     &          &       &             \cr
\ph0 &       &       &     &          &       &             \cr
\ph0 &       &       &     &          &       &             \cr}
		}
where the $\rho$ block is $n_c \times n_c$, implying
		\eqn\Nbb{
N = \pmatrix{
\ph0 &\ph0 &\ph0 &\ph0 &\lambda_1 &       &             \cr
     &     &     &     &          &\ddots &             \cr
     &     &     &     &          &       &\lambda_\tnc \cr
     &     &     &     &0         &\cdots &0            \cr
     &     &     &     &-\rho     &       &             \cr
     &     &     &     &          &\ddots &             \cr
     &     &     &     &          &       &-\rho        \cr}.
		}
But it is now straight-forward to check using these solutions that
$(*B)\,\tB = b\,(*\tb) = (-\rho)^{n_c - \tnc}\, *(N^\tnc) = *(M^{n_c})$,
showing that \mbconstri\ is satisfied.  When both $b$ and $\tb$ vanish,
either Tr$N = 0$ or rank$(N) < \tnc$.  In either case
\bbrconstriv\ then implies rank$(M) < n_c$, so that \mbconstri\ is
again satisfied.

So, we have shown that the Higgs branches of the
IR--effective theory \macbary\ are a subset of those of the microscopic
theory.  Furthermore, we have seen explicitly that the baryonic branch
is a solution to both sets of constraints.  The difference between the
two sets is in the non-baryonic branches.  In particular, on the
non-baryonic branches (since Tr$N = 0$) we have from
\bbrconstriv\ that
		\eqn\Mnbb{
N = M = \pmatrix{
\ph0 &\ph0 &\lambda_1 &       &          \cr
     &     &          &\ddots &          \cr
     &     &          &       &\lambda_r \cr
\ph0 &     &          &       &          \cr
\ph0 &     &          &       &          \cr}, \qquad
r \le [n_f/2].  } 
However, the ``extra'' constraint \nrankconstr\ implies that of the
branches \Mnbb\ only those with $r \le \tnc$ survive.  Thus the
IR--effective theory at the root of the baryonic branch includes a
subset of the non-baryonic branches of the microscopic theory.

\subsec{$N{=}2$ Duality and Flowing Down in Flavors}

In the previous subsection we determined the IR--effective theory at the
root of the baryonic branch using the unbroken $\bZ_{n_c-\tnc}$
symmetry to argue that there should be an unbroken $SU(\tnc)\times
U(1)^{n_c-\tnc}$ gauge group, and demanding singlets charged under the
$U(1)$'s so the Higgs branch would exist, leading to the quantum numbers 
\macbary.  We then checked that \macbary\ is indeed the effective theory
at the baryonic root by showing that its Higgs branches are 
the baryonic Higgs branch as well as
$\tnc$ of the $[n_f/2]$ non-baryonic branches.

In this subsection we will perform a further check which illuminates the
connection between the effective theory at the baryonic root and a
conjectured S--duality of a finite $N{=}2$ theory.  In particular,
we show that \macbary\ can be obtained by starting with the
scale-invariant theory with $2n_c$ flavors and flowing down to $n_f$
flavors by giving bare masses to $n_c {-} \tnc$ quarks.  We determine
the form of this bare mass term in the IR--effective theory using the
conjectured S-duality of the finite $N{=}2$ theory.

Consider the scale-invariant theory with $2n_c$ flavors, coupling
$\tau$, and bare masses $m_i$.  The exact solution for this theory
\APS\ shows it has a symmetry implying an equivalent description in
terms of a dual theory with coupling and masses replaced by 
		\eqn\Edual{\eqalign{ 
\tau &\rightarrow \til\tau = -1/\tau ,\cr 
m_i  &\rightarrow \til m_i = m_i - 2m_S ,
		}}
where $m_S$ is the flavor-scalar mass defined by
		\eqn\Emass{ 
m_S \equiv {1\over n_f} \sum_{i=1}^{n_f} m_i .
		}
Note that under the duality transformation $\til m_S = -m_S$ while
the adjoint mass $\til m_{Ai} = \til m_i {-} \til m_S = m_i {-}  m_S=
m_{Ai}$ is invariant.  This fact has a natural interpretation in terms
of the non-renormalization theorem of section 3.  Since we think of
$m_i$ as background vector superfields, their couplings are constrained
by gauge invariance and the coefficients in the superpotential \Niii\ is
the corresponding charge.  For $m_S$ that charge is the baryon number.
Therefore, the transformation \Edual\ means that the baryon number of
the magnetic quarks is opposite to that of the electric quarks.

We flow down to the theory with $n_f < 2n_c$ by turning on $n_c {-}
\tnc$ quark masses of order $M$:
		\eqn\Eemass{
m_i = (\ \underbrace{\ M x_i\ }_{n_c-\tnc}\ ;\ \underbrace{\ \,0\,\ 
}_{n_f}\ ).
		}
The resulting theory should not depend on the way we decouple the
$n_c{-}\tnc$ flavors, so the $x_i$ should be taken to be arbitrary
numbers satisfying $|x_i|\roughly{>} 1$.  The massive quarks decouple
in the 
limit $q\equiv e^{i\pi\tau} \rightarrow 0$ and $M\rightarrow\infty$
keeping $\Lambda \equiv q^{1/(n_c-\tnc)}M$ fixed.  The dual of this
theory is at very strong coupling with masses $\til m_i \sim M$:
		\eqn\Edmass{
\til m_i=\Bigl(\ \underbrace{Mx_i-{M\over n_c}\sum x}_{n_c-\tnc}\ ;\
\underbrace{-{M\over n_c}\sum x}_{n_f}\ \Bigr).
		}

Our strategy for determining the spectrum of the effective theory at
the root of the baryonic branch will be to study this dual theory at
weak coupling instead of strong coupling.  The point is that since this
theory is IR--free, it will exist along whole submanifolds in parameter
space ($\til m_i$, $\til\tau$) which can plausibly be followed out to
weak coupling.

A unique point $\phi$ on the Coulomb branch of the dual theory is
determined by the requirement of IR--freedom and the existence of a
purely hypermultiplet Higgs branch.  Since for generic $x_i$ the first
$n_c{-}\tnc$ masses in \Edmass\ are all different, by tuning $\phi$
each can contribute at most one massless flavor which, to be IR--free,
must be charged only under $U(1)$ factors.  In order to have a purely
hypermultiplet Higgs branch, however, we must have at least as many
massless singlets as $U(1)$ factors.  These conditions determine the
diagonal $\phi$ vev
		\eqn\Ecbpoint{
\phi=\Bigl(\ \underbrace{-Mx_i+{M\over n_c}\sum x}_{n_c-\tnc}\ ;
\ \underbrace{{M\over n_c}\sum x}_\tnc\ \Bigr),
		}
giving an $SU(\tnc)\times U(1)^{n_c-\tnc}$ effective theory with $n_f$
massless squarks, and $n_c{-}\tnc$ massless singlets each of which is
charged under a single $U(1)$ factor.  Normalizing these $U(1)$'s so
that each singlet has charge $-1$, the squarks must then have charge
$1/\tnc$  under each of the $U(1)$ factors, as follows from the
tracelessness of each factor inherited from its embedding in the
original $SU(n_c)$ group.  We thus recover the baryonic root effective
theory \macbary.

We can check the baryon number of the dual squarks in \macbary\ as well, 
using their coupling to the flavor-scalar mass.  Consider adding in 
additional small bare masses $m'_i$ for the squarks in the microscopic
theory.  The bare masses $m_i$, their duals $\til m_i$, and the vev
for $\phi$ are then
		\eqn\Evevs{\eqalign{
m_i &= (\ \underbrace{\ M x_i\ }_{n_c-\tnc}\ ;\ \underbrace{\  
m'_\hi\ }_{n_f}\ ), \cr
\til m_i &= \Bigl(\ \underbrace{M x_i -{M\over n_c} \sum x - {1\over 
n_c}\sum m'}_{n_c-\tnc}\ ;\ \underbrace{m'_\hi -{M\over n_c} \sum x -
{1\over n_c}\sum m'}_{n_f} \ \Bigr), \cr
\phi &= \Bigl(\ \underbrace{-M x_i + {M\over n_c} \sum x + {1\over n_c}
\sum m'}_{n_c-\tnc}\ ; \ \underbrace{{M\over n_c} \sum x - {n_c-\tnc\over 
n_c\tnc}\sum m'}_{\tnc}\ \Bigr),
		}}
with $|M|\gg |m'_i|$.  On the Coulomb branch of the dual theory near
$\phi$ the effective theory is $SU(\tnc)\times U(1)^{n_c{-}\tnc}$ with
$n_f$ squarks with masses
		\eqn\Egenmassdual{
\til m'_i = m'_i - {1\over\tnc}\sum m',
		}
and we tuned $\phi$ so the $n_c{-}\tnc$ singlets are exactly massless.
(The fact that we can shift these masses away shows that they are not
true parameters describing the vacuum in question.)  The expression for
the quark masses in the dual theory \Egenmassdual\ implies the duality
transformation on the flavor-scalar mass
		\eqn\ENACdual{
 m_S \rightarrow \til m_S =  -{n_c\over\tnc}  m_S .
		}
Recalling our interpretation of the masses as the vevs of vector
multiplets gauging the flavor symmetries, it follows that the
coefficient of the scalar mass in the superpotential should be
identified with the baryon number of the squarks.  In the microscopic
theory the singlet mass couples as
		\eqn\EWelec{
{\cal W}_{\rm micro}/\sqrt2 = +  m_S\, {\rm tr}Q{\cdot}\til Q ,
		}
normalizing the baryon number of the squarks to $1$.  In the dual
theory, we have
		\eqn\EWmagn{
{\cal W}_{\rm dual}/\sqrt2 = 
+ \til m_S\, {\rm tr}q{\cdot}\til q = -{n_c\over
\tnc}  m_S\, {\rm tr}q{\cdot}\til q ,
		}
implying the dual squarks have baryon number $-n_c/\tnc$.

		\nref\GMS{B.R. Greene, D.R. Morrison, and A. Strominger,
	\hepth{9504145}, \NPB{451}{1995}{109}.
		} 
We could also have checked the assignment of charges \macbary\ by
flowing down one flavor at a time using \Egenmassdual.
It is worth pointing out that one finds in this way the spectrum for
the extreme $n_f=n_c$ case (which is not covered in \macbary):
\foot{It may be interesting to note that these quantum
numbers (for $n_c=16$) correspond to the massless spectrum at the
conifold point studied in \GMS. Is there a non-Abelian ({\it e.g.} a
heterotic dual) theory whose long-distance behavior leads to this
spectrum at that point?}  
		\eqn\macbaryii{\matrix{
& U(1)_1 &\times& \cdots &\times& U(1)_{n_c-1} &\times& U(1)_B \cr
e_0       & 1     && \cdots && 1      &&-n_c    \cr
e_1       &-1     && \cdots && 0      && 0      \cr
\vdots    &\vdots && \ddots && \vdots && \vdots \cr
e_{n_c-1} & 0     && \cdots &&-1      && 0      \cr
		}}

\newsec{Quantum Coulomb Branch}

Though the arguments of the last section determined the low energy physics
on the Coulomb branch at the roots of the Higgs branches, they did not
determine the positions of these roots on the Coulomb branch.
In this section we will locate these points using the exact solution for
the effective theory on the Coulomb branch.  While, on the one hand,
locating these points can be viewed as a check on the arguments of the
last section, on the other hand, given that those arguments only
required semi-classical physics, this section may be better viewed as
a check on the exact solutions.

\subsec{Review of the Exact Solution}

The generic vacuum on the Coulomb branch is a $U(1)^{n_c-1}$ pure Abelian
gauge theory characterized by an effective coupling $\tau_{ij}$ between
the $i$th and $j$th $U(1)$ factors.  Due to the ambiguity of
electric-magnetic duality rotations in each $U(1)$ factor, as well as
changes of basis among the $U(1)$'s, the $\tau_{ij}$ form a section of an
$Sp(n_c{-}1,\bZ)$ bundle over the Coulomb branch \SWi.

Locally the Coulomb branch is $n_c {-} 1$ complex dimensional,
corresponding to the vevs of the $n_c {-} 1$ $U(1)$ vector multiplets.
Globally, it is found in \refs{\AF-\HO} that the Coulomb branch can be
characterized by $n_c$ complex numbers
$\Phi=(\phi_1,\ldots,\phi_{n_c})$ (up to permutations) whose sum
vanishes.  At weak coupling these coordinates are identified with the
coordinates of the adjoint scalar vev in the Cartan subalgebra of
$SU(n_c)$.

An explicit description of the Coulomb branch was found in \refs{\AF-\HO}
by associating to each point of the Coulomb branch a genus $n_c {-} 1$
Riemann surface whose complex structure is the low energy coupling
$\tau_{ij}$.  This family of Riemann surfaces is described by complex
curves $y^2 = P_{2n_c}(x)$ where $P_{2n_c}$ is a polynomial of degree
$2n_c$ in $x$ whose coefficients are polynomials in the moduli
$\phi_a$ and the parameters---the bare quark masses $m_i$, and the
strong-coupling scale $\Lambda$ or bare coupling $\tau$ (in the latter
case the dependence is through $\theta$ functions).  Such a double
cover of the complex $x$-plane branched over $2n_c$ points describes a
genus $n_c {-} 1$ hyperelliptic Riemann surface.  When two or more of the
branch points (the zeros of $P_{2n_c}$) collide as we vary the moduli
or parameters, the Riemann surface degenerates.  Such a singularity in the
effective action corresponds to additional $N{=}2$ multiplets becoming
massless.

The mass of any BPS saturated state in the theory is given by the period
of a certain one-form around a cycle on the Riemann surface \SWi.  The
homology class of the cycle encodes the $U(1)$ charges of the state.
States corresponding to non-intersecting cycles are ``mutually local''
in the sense that they can be taken to be simultaneously solely
electrically charged under the $U(1)$ factors.  Thus, when only two branch
points collide the cycle encircling the two points vanishes,
corresponding to vanishing mass for hypermultiplet states with
$U(1)$ charges proportional to the homology class of the vanishing
cycle.    Near such a point, the 
curve will have the form $y^2 = P_{2n_c-2}(x) \cdot (x {-} x_1)^2$ for
some 
$P_{2n_c-2}$ and $x_1$.  The number of hypermultiplets (weighted by
their charges) is determined by the monodromy of the periods as we
move $\phi$ about the singular point.  In the generic case in which
there is precisely one massless hypermultiplet, we will refer to such
a curve as having a single 
hypermultiplet singularity.  Similarly, if $n_s$ independent pairs of
zeros of $P_{2n_c}$ collide, the curve is $y^2=P_{2n_c-2n_s} \cdot
\prod_{a=1}^{n_s}(x {-} x_a)^2$, and has generically an $n_s$
(mutually local) hypermultiplet singularity.

When more than two zeros collide at a point, more complicated sets of
states become massless.  In some cases the resulting effective theories
are non-trivial fixed points \refs{\AD,\APSW}.  Others may be IR--free or
scale-invariant non-Abelian Coulomb points.  We will now develop a
``dictionary'' for identifying the latter type of singularity from the
form of the curve.  (The beginnings of such a dictionary for the
non-trivial fixed points appears in \APSW.)

The curve for the $n_f{=}2n_c$ scale-invariant theory is \APS
		\eqn\Escalecurv{
y^2 = \prod_{a=1}^{n_c}(x-\phi_a)^2 + h(h+2) \prod_{i=1}^{n_f} (x + h
 m_S + m_i), \qquad n_f = 2n_c
		}
where $ m_S = (1/n_f)\sum m_i$ is the singlet mass, and $h(q) = 32 q +
{\cal O}(q^2)$ where $q=e^{i\pi\tau}$, is a specific modular function
of $\tau$: $h(\tau) = 2\theta_1^4 / (\theta_2^4 {-} \theta_1^4)$.
(We define the $\theta_i(\tau)$ as in \SWii.) The duality symmetry of
the curve under the transformation \Edual\ described earlier follows
{}from the modular transformation property $h \rightarrow -(h{+}2)$
under $\tau \rightarrow -1/\tau$ \APS.  

One can deduce the $n_f > 2n_c$ curves by starting with the
scale-invariant theory with $n_f/2$ colors and breaking the gauge group
at a scale $M$ down to $SU(n_c)$ on the Coulomb branch at weak coupling
($h {\sim} 0$) with $\Lambda^{2n_c - n_f} = 16q M^{2n_c - n_f}$
fixed.  (The factor of $16$ is for later convenience.)  This
corresponds to setting $|\phi_a| \ll |\phi_k| \sim M$ for $a = 1,
\ldots, n_c$ and $k = n_c{+}1, \ldots, n_f/2$ with $\sum_a \phi_a =0$.
Such a procedure makes sense since the $SU(n_c)$ theory with $n_f$
flavors is IR--free, so for any $\Lambda$ the region of the Coulomb
branch in question is weakly coupled.  In the limit $q \rightarrow 0$
we find the curve
		\eqn\IRfreecrv{
y^2 = \prod_{a=1}^{n_c}(x - \phi_a)^2 + 4\Lambda^{2n_c-n_f}
\prod_{i=1}^{n_f} (x + m_i), \qquad n_f > 2n_c,
		}
for $|x|, |\phi_a|, |m_i| \ll M$.  $M$ is the scale at which the
description of the physics by the curve \IRfreecrv\ must break down (it
becomes sensitive to the physics that regularizes it in the UV at
scales above $M$).  $\Lambda$ is the scale where this IR--free theory
becomes strongly coupled.  Note that at weak coupling $M \ll
\Lambda$, and the cutoff scale $M$ cannot be determined from the curve
\IRfreecrv\ alone.  These considerations are reflected in the fact that
although \IRfreecrv\ is supposed to describe the Coulomb branch of an
$SU(n_c)$ theory, it involves a polynomial in $x$ of degree $n_f >
2n_c$.  But the $2n_c {-} n_f$ ``extra'' zeros of the right-hand side
of \IRfreecrv\ are at $x \sim \Lambda$, and so are of no consequence in
the region of validity $|x| \ll M \ll \Lambda$ of \IRfreecrv.

The curves for the asymptotically-free theories with $n_f < 2n_c$ are
obtained from the curve for the scale-invariant theory by integrating
out some of the flavors.  Thus, we can determine the curve for $n_f=
2n_c {-} 1$ by taking $q {\rightarrow} 0$ and one of the masses, say
$m_{2n_c}$, large such that $\Lambda = 16 qm_{2n_c}$ stays fixed, thus
decoupling that flavor.  $\Lambda$ becomes the strong-coupling scale of
the resulting AF theory.  The curve becomes for $|x|\ll|m_{2n_c}|$
		\eqn\Eimasscrv{
y^2 = \prod_{a=1}^{n_c}(x-\phi_a)^2 + 4\Lambda\prod_{i=1}^{n_f} \left(x
+ m_i + {\Lambda\over n_c} \right), \qquad n_f = 2n_c -1 ,
		}
thus describing the Coulomb branch of the $n_f = 2n_c {-}1$ theory for
all values of $x$, $\phi_a$, and $m_i$.  For two or more large masses,
say $m_i{\rightarrow}\infty$ for $n_f{+}1 \le i \le 2n_c$, take
$q{\rightarrow}0$ such that $\Lambda^{2n_c-n_f}= 16 q
\prod_{i=n_f+1}^{2n_c} m_i$ stays fixed.  The curve becomes for $|x|\ll
\min_{n_f+1\le i\le 2n_c}\{|m_i|\}$
		\eqn\Eiimasscrv{
y^2 = \prod_{a=1}^{n_c}(x-\phi_a)^2 + 4\Lambda^{2n_c-n_f}
\prod_{i=1}^{n_f}(x+m_i) , \qquad n_f \le 2n_c -2,
		}
describing the Coulomb branch of the $n_f \le 2n_c {-}2$ theories.

\subsec{The Non-Baryonic Roots}

{}From the discussion of section 4.1, the non-baryonic roots have
unbroken $SU(r) \times U(1)^{n_c-r}$ gauge symmetry for $r \le
[n_f/2]$, with $n_f$ massless quarks.  This suggests we look at a
submanifold of the Coulomb branch of the form
		\eqn\nbbrphicoords{
\Phi_{\rm nbb} = (0,\ldots,0,\varphi_1,\ldots,\varphi_{n_c-r}), \qquad
\sum \varphi_a = 0, \qquad |\varphi_a| \sim M.
		}
The curve with zero masses for $n_f \le 2n_c{-}2$ is $y^2 =
\prod_a^{n_c} (x {-} \phi_a)^2 + 4 \Lambda^{2n_c-n_f}x^{n_f}$.  For
$\Phi = \Phi_{\rm nbb} {+} \delta\Phi$ near $\Phi_{\rm nbb}$,
$\delta\Phi = (\phi_1, \ldots, \phi_r, 0, \ldots, 0)$, $\sum \phi_a
=0$, and $|\phi_a|, |x| \ll M$, the curve becomes approximately
		\eqn\Ecrvii{
y^2 = \prod_a^r (x-\phi_a)^2 M^{2(n_c-r)} -
4\Lambda^{2n_c-n_f}x^{n_f} ,
		}
which we recognize as the curve for $SU(r)$ with $n_f$ flavors,
confirming \nbbrphicoords\ as the coordinates of the non-baryonic
roots.  (For $n_f = 2n_c {-} 1$ the coordinates have to be shifted to
$\Phi_{\rm nbb} = (0,\ldots,0,\varphi_1 {-} {1\over
r}\Lambda,\ldots,\varphi_{n_c-r} {-} {1\over r}\Lambda) - {1\over
n_c}\Lambda (1, \ldots, 1)$.  Then, for $\til x \equiv x {+} {1\over
n_c} \Lambda \ll \Lambda$, \Eimasscrv\ becomes approximately \Ecrvii.)

		\nref\DS{M.R. Douglas and S.H. Shenker,
	\hepth{9503163}, \NPB{447}{1995}{271}.
		}
Now we set $\delta\Phi=0$ and look for points on the $\Phi_{\rm nbb}$
submanifold for which we get the maximal number of mutually local
massless monopoles.  The curve becomes
		\eqn\Eci{
y^2 = x^{2r} \left\{ \prod_a^{n_c-r} (x-\varphi_a)^2 + 
4\Lambda^{2n_c-n_f} x^{n_f-2r}\right\}.
		}
Since we have $n_c {-} r {-} 1$ parameters $\varphi_a$ at our disposal,
we can (generically) tune $2(n_c {-} r {-} 1)$ zeros of \Eci\ to double
up.  By choosing a basis of contours on the $x$-plane with
$n_c{-}r{-}1$ which each loop around one of the doubled zeroes and the
rest which loop in various ways around the $2r$ zeros corresponding to
the $SU(r)$ factor, we find (generically) the non-baryonic root charges
given in \Enonbary.  It is difficult to determine the actual values of
$\varphi_a$ for which this coincidence of zeros is realized.
{}From the discrete $\bZ_{2n_c-n_f}$ symmetry of the curve \Eci, it
follows that there will generically be $2n_c {-} n_f$ such points.
When $r = n_f/2$, the curve in braces in \Eci\ describes the pure
$SU(n_c {-} n_f/2)$ Yang-Mills theory, and explicit coordinates for its
multimonopole points have been computed in \DS.  When $r \leq n_f{-}n_c$,
there is in fact a point at which the discrete symmetry is unbroken
and there are $n_c{-}r$ massless hypermultiplets.  This point is the
baryonic root found in the next subsection.  For $r<n_f{-}n_c$ there are
also points as described above with $n_c{-}r{-}1$ massless
hypermultiplets 
and a broken discrete symmetry.  Thus we see (as discussed following
\Mnbb ) that quantum mechanically the nonbaryonic branches with $r\leq
n_f{-}n_c$ intersect the baryonic branch.

\subsec{The Baryonic Root}

{}From the discussion in section 4.1, we expect the effective theory at
the baryonic root to have an $SU(n_f{-}n_c)\times U(1)^{n_c-\tnc}$ gauge
symmetry with $n_f$ massless quarks, and to be invariant under the
$\bZ_{n_c-\tnc}$ discrete symmetry on the Coulomb branch.  Thus (for
$n_f \le 2n_c{-}2$) we take
		\eqn\bbrphicoords{
\Phi_{\rm bb} = (0, \ldots 0, \varphi\omega, \ldots,
\varphi\omega^{n_c-\tnc}),
		}
where $\omega = {\rm exp}\{2\pi i/(n_c{-}\tnc)\}$.  Indeed, for $\Phi =
\Phi_{\rm bb} {+} \delta\Phi$ near $\Phi_{\rm nbb}$, $\delta\Phi =
(\phi_1, \ldots, \phi_\tnc, 0, \ldots, 0)$, $\sum \phi_a =0$, and
$|\phi_a|, |x| \ll \varphi$, the curve becomes approximately
		\eqn\Ecrvii{
y^2 = \prod_a^\tnc (x-\phi_a)^2 \varphi^{2(n_c-\tnc)} +
4\Lambda^{n_c-\tnc} x^{n_f} ,
		}
which we recognize as the curve for $SU(\tnc)$ with $n_f$ massless
flavors.

We now determine $\varphi$ such that $n_c {-} \tnc$ mutually local
hypermultiplets become massless.  Setting $\delta\Phi=0$, the curve
becomes 
		\eqn\Ecrviii{
y^2 = x^{2\tnc} \left[ (x^{n_c-\tnc} - \varphi^{n_c-\tnc})^2 + 
4\Lambda^{n_c-\tnc} x^{n_c-\tnc} \right].
		}
At the point
		\eqn\ENACpnt{
\varphi = \Lambda 
		}
the right hand side of \Ecrviii\  becomes a perfect square,
giving the right singularity structure to describe $n_c{-}\tnc$
massless hypermultiplets.  This is thus the point we want.\foot{We can
also check that there are no such multimonopole points on the larger
$\bZ_{n_c - \tnc}$ invariant submanifolds for $n_f \ge {3\over2}n_c$
mentioned in the footnote after Eq.\ \bbrcoord.  On this submanifold the
curve becomes $y^2 = x^{2n_c - 2k\nu} \cdot (\prod_{i=1}^k (x^\nu -
\varphi_i^\nu)^2 + 4\Lambda^\nu x^{(2k-1)\nu})$, where $\nu \equiv n_c -
\tnc$.  It is easy to show that the right hand side becomes a perfect
square only for all but one $\varphi_i =0$, which restricts us back to the
submanifold \bbrphicoords.} It is easy to check that there is indeed
precisely one hypermultiplet of each set of charges, by computing the
monodromy. (For $n_f=2n_c{-}1$
we have to shift $\Phi_{\rm bb}$ to $\Phi_{\rm bb} = (0, \ldots, 0,
n_c \Lambda) - \Lambda (1, \ldots,1)$.)

Furthermore, if we take a basis of cycles on the $x$-plane with $n_c
{-} \tnc$ cycles $\alpha_i$ which each encircle one of the $n_c {-}
\tnc$ pairs of zeros of \Ecrviii, and the rest of the cycles looping in
various ways around the $2\tnc$ zeros associated with the $SU(\tnc)$
factor, we see that the singlets are each charged under only one $U(1)$
factor.  For small $\phi_a$ in \Ecrvii\ the $2\tnc$ degenerate zeros
split.  Consider a set of $\tnc$ non-intersecting contours $\gamma_a$
each looping once around one pair of these slightly split zeros.  Thus
each $\gamma_a$ corresponds to a light quark.  A contour deformation on
the $x$-plane shows that the set $\{\alpha_i,\gamma_a\}$ of contours
satisfy one relation $\sum_i \alpha_i + \sum_a \gamma_a = 0$.  Since
the $\gamma_a$ do not sum to zero, they do not correspond to a basis of
$U(1)$'s in the Cartan subalgebra of the $SU(\tnc)$ factor.  We can
define an alternative basis of cycles $\{\alpha_i, \beta_a\}$ with
$\beta_a = \gamma_a + {1\over\tilde n_c}\sum_i \alpha_i$ which does
have this property.  In this basis each quark contour $\gamma_a$ has a
$-{1\over\tilde n_c} \alpha_i$ piece, and so we conclude that the
quarks are each charged under all the $U(1)$ factors with $-1/\tnc$ of
the corresponding singlet charge.  Thus we have reproduced the quantum
numbers for the baryonic root derived by other arguments in section 4.

\newsec{Breaking to $N{=}1$ Supersymmetry}

In this section we break to $N{=}1$ supersymmetry by turning on a
bare mass $\mu$ for the adjoint superfield $\Phi$.  In the microscopic
theory in the limit $\mu {\rightarrow} \infty$ this leads to $N{=}1$
$SU(n_c)$ super--QCD.  By performing this breaking in the macroscopic
effective theories at the roots of the Higgs branches we find instead
$SU(\tnc)$ super--QCD with some extra gauge singlets.  This is the
dual formulation of $N{=}1$ QCD suggested in \Sei.

\subsec{Breaking in the Microscopic Theory}

Since $\Phi$ is part of the $N{=}2$ vector multiplet, giving it a mass
explicitly breaks $N{=}2$ supersymmetry.  In the microscopic theory,
this corresponds to an $N{=}1$ theory with a superpotential
		\eqn\Nisupotl{
{\cal W} = \sqrt2\, {\rm tr}(Q\Phi \tQ) + {\mu \over 2} {\rm tr}(\Phi^2) .
		}
For $\mu\gg\Lambda$ we can integrate $\Phi$ out in a weak-coupling
approximation, obtaining an effective quartic superpotential
		\eqn\Nieffsp{
{\cal W}_\prime =  -{1\over \mu} \left( {\rm tr}(Q\tQ Q\tQ) - {1\over
n_c} {\rm tr}(Q\tQ) {\rm tr}(Q\tQ) \right).
		}
In the limit $\mu {\rightarrow} \infty$ this superpotential becomes
negligible, and we find $N{=}1$ $SU(n_c)$ super--QCD with $n_f$ flavors
and no superpotential.

Note that if the strong coupling scale of the $N{=}2$ theory is
$\Lambda$, then by a one-loop matching, the scale of the $N{=}1$
theory will be $\Lambda_1^{3n_c - n_f} = \mu^{n_c} \Lambda^{2n_c -n_f}$.
The appropriate scaling limit will send $\mu \rightarrow \infty$ and 
$\Lambda \rightarrow 0$ keeping $\Lambda_1$ fixed.  The model is
described by the $N{=}1$ super--QCD theory on scales between $\mu$ and
$\Lambda_1$, below which the strongly-coupled dynamics of the $N{=}1$
theory takes over. 

\subsec{Breaking in the Low-Energy Effective Theories}

We can also study the breaking to $N{=}1$ by beginning with
$\mu\ll\Lambda$.  In this case we should study the low-energy $N{=}2$
theory obtained in the previous sections and the effects of $\mu$ in
this theory.  $N{=}1$ supersymmetry prevents a phase transition as we
vary $\mu$, hence we should obtain the same result as that obtained
for $\mu\gg\Lambda$ in the previous subsection (see Fig. 3). 

It is easy to see that generic vacua of the $N{=}2$ theory are lifted
by nonzero $\mu$; we will show that the baryonic root, as well as the
special points we have found on the nonbaryonic roots, are not.  There
may be other points that survive the breaking, but the light fields at
these points will not include non-Abelian gauge multiplets.  We thus
study the effects of the breaking to $N{=}1$ in the effective theories
at the roots of the Higgs branches.  We saw in section 4 that
these effective theories have unbroken gauge groups of the form $SU(r)
\times U(1)^{n_c-r}$.  Let $\phi$ denote the adjoint scalar in the
$SU(r)$ factor, and $\psi_k$ the adjoint scalars for each of the $U(1)$
factors.  Then a microscopic mass term $(\mu/2){\rm tr}\Phi^2$ becomes
$\mu(\Lambda\sum_i x_i \psi_i + {1\over2}{\rm tr}\phi^2 + \ldots)$,
where the dots denote higher-order terms, and $x_i$ are dimensionless
numbers.  From
the $\Phi$ vevs breaking $SU(n_c) \rightarrow SU(r) \times
U(1)^{n_c-r}$ found in  section 5, we see that all $x_i \sim 1$.  

Now let us examine what happens to the effective theory at the roots of
the non-baryonic branches when we turn on such a mass term.  The first
thing to note is that at any point on the non-baryonic root for which
there are fewer than $n_c{-} r{-} 1$ massless singlets, $e_k$, charged
under the $U(1)$'s, then the $N{=}2$ vacuum is lifted.  This can be
seen as follows.  If there are $n_s$ singlets with $n_s < n_c{-} r{-}
1$, a basis of the $U(1)$'s can be chosen to diagonalize the charges of
the singlets and the quarks, and the superpotential becomes
		\eqn\Ninbbrsp{
{\cal W}_{\rm nbb} = \sqrt2\, {\rm tr}(q\phi\tq) + 
\sqrt2\,\psi_0 {\rm tr}(q\tq) +
\sqrt2\sum_{k=1}^{n_s} \psi_k e_k \til e_k + \mu\Bigl( \Lambda
\sum_{i=0}^{n_c-r-1} x_i \psi_i + {1\over 2} {\rm tr} \phi^2 \Bigr).
		}
The $F$-term equations following from taking derivatives with respect
to the $\psi_i$ then have no solution.

Therefore only the special vacua \Enonbary\ on the non-baryonic roots
with $n_s= n_c{-} r{-}1$ lead to $N{=}1$ vacua.  Then the $\psi_i$
$F$-term equations imply $e_k\til e_k\neq 0$, while the $e_k$ equations
imply that all $\psi_i=0$ except $\psi_0$.  Thus when $\mu \neq 0$
these fields can be integrated-out, leaving the effective
superpotential
		\eqn\Ninbbreff{
{\cal W}_{\rm nbb}^\prime = \sqrt2\,{\rm tr}(q\phi\tq) + 
{\mu\over2}{\rm tr} 
\phi^2 + \psi_0 \left(\sqrt2\, {\rm tr}(q\tq) + \mu\Lambda \right),
		}
for an $N{=}1$ $SU(r)\times U(1)$ super--QCD, with $r \le [n_f/2]$.
The $q$ and $\phi$ $F$-term equations then imply a trade-off between
the rank of the unbroken gauge group and the number of massless
singlets.  The general solutions up to color and flavor rotations
are as follows. 
		\eqn\EFsoln{\eqalign{
\psi_0 &= {\ell{-}r\over\ell r} \Lambda,\qquad 
\ell\in\{0,1,\ldots,r\},\cr
\phi &= {\Lambda\over \ell r} {\rm diag}
(\underbrace{r{-}\ell,\ldots,r{-}\ell}_{\ell}, 
\underbrace{-\ell,\ldots,-\ell}_{r-\ell}),\cr
q &= \pmatrix{
\kappa_1    &&& \ph{0_1}    &&& \ph{0}&\cr 
&\ddots      && &\ph{\ddots} && &\ph{0}\cr
&&\kappa_\ell & &&\ph{0_\ell} & &      \cr
&&&             &&&             &\ph{0}\cr},
\qquad \kappa_a\in\bR^+ ,\cr 
\t\tq &= \pmatrix{
\tk_1       &&& \lambda_1    &&& \ph{0}&\cr 
&\ddots      && &\ddots       && &\ph{0}\cr
&&\tk_\ell    & &&\lambda_\ell & &\ph{0}\cr 
&&&             &&&              &\ph{0}\cr},
\qquad \lambda_a\in\bR^+ ,\cr
		}}
where $\kappa_a\tk_a = -\mu\Lambda/\ell$, independent of $a$, and
$\lambda^2_a = \kappa_a^2 {-} \tk_a^2$.  This classical moduli space of
solutions all have unbroken $SU(r{-}\ell)$ gauge symmetry with
$\ell(n_f {-} \ell)$ massless singlets, and no light charged matter.
The corresponding quantum theories have no IR gauge group, since
$N{=}1$ Yang-Mills theory is known to be confining.  The massless
chiral multiplets along these flat directions---the meson fields $N^i_j
= q^i \tq_j$---should be identified with the meson fields $M$ in the dual
$N{=}1$ theory of \Sei , since they are gauge singlets, have zero
baryon number, and transform in the adjoint plus singlet
representations of $SU(n_f)$ 
flavor.  Furthermore, since rank$(N) \le [n_f/2]$, $N$ cannot be
identified with the meson field $\tilde M^i_j$ bilinear in the dual
$N{=}1$ quarks since rank$(\tilde M) \le \tnc < [n_f/2]$.  Therefore
$N$ must be (part 
of) the ``extra'' singlet degrees of freedom $M$ found in the $N{=}1$
dual of \Sei.

Now let us examine the breaking of the effective theory at the baryonic
root.  In this case, from \macbary, the superpotential is
		\eqn\Nibbrsp{
{\cal W}_{\rm bb} = \sqrt2\,\tr(q\phi\tq) + {\sqrt2\over\tnc} \tr(q\tq) 
\Bigl(\sum_{k=1}^{n_c-\tnc} \psi_k\Bigr) - \sqrt2
\sum_{k=1}^{n_c-\tnc} \psi_k e_k \til e_k +
\mu\Bigl(\sum_{i=1}^{n_c-\tnc} \Lambda_i
\psi_i + {1\over 2} {\rm tr} \phi^2 \Bigr),
		}
for an $N{=}1$ $SU(\tnc)$ theory.  In addition to some flat directions
similar to those in \EFsoln\ in which the quarks in the fundamental of
the unbroken gauge group are all lifted and there are various massless
mesons, in this case there are also vacua where the $e_k$ get vevs,
Higgsing all the $U(1)$ factors.  Integrating out these massive fields,
we find the effective superpotential
		\eqn\Nibbrspi{
{\cal W}_{\rm bb}^\prime = \sqrt2 \tr(q\phi\tq)+ 
{\mu\over2}{\rm tr}\phi^2, 
		}
describing $SU(\tnc)$ super--QCD with $n_f$ flavors in the limit $\mu
{\rightarrow} \infty$, thus showing the origin of the ``magnetic''
gluons and quarks of the dual $N{=}1$ theory of \Sei.

\bigskip
\centerline{{\bf Acknowledgments}}

It is a pleasure to thank T. Banks, K. Intriligator, S.  Shenker,
A. Schwimmer, E. Witten and especially A. Shapere for helpful
discussions and comments. This work was supported in part by 
DOE grant \#DE-FG05-90ER40559 and by the Center for Basic Interactions.

\listrefs
\end